\pdfoutput=1

\PassOptionsToPackage{dvipsnames}{xcolor}
\documentclass[acmsmall, screen]{acmart}

\usepackage{xcolor}
\usepackage{xspace}
\usepackage{booktabs, multirow}
\usepackage{soul}
\usepackage{listings}
\usepackage{subcaption}
\usepackage{wrapfig}
\usepackage[ruled, vlined, linesnumbered, commentsnumbered, longend]{algorithm2e}
\usepackage[normalem]{ulem}

\newcommand{\eg}{\emph{e.g.,}\xspace}
\newcommand{\ie}{\emph{i.e.,}\xspace}

\definecolor{applegreen}{rgb}{0.55, 0.71, 0.0}

\newcommand{\revision}[1]{#1}

\newcommand{\verus}{{Verus}\xspace}

\newcommand{\lynette}{{Lynette}\xspace}
\newcommand{\verusbench}{{Verus-Bench}\xspace}

\newcommand{\mbpp}{{MBPP}\xspace}
\newcommand{\diffy}{{Diffy}\xspace}
\newcommand{\clover}{{CloverBench}\xspace}
\newcommand{\misc}{{Misc}\xspace}

\newcommand{\inferenceData}{{Inference Dataset}\xspace}
\newcommand{\refinementData}{{Refinement Dataset}\xspace}

\newcommand{\sampleData}{{Sampled-dataset}\xspace}

\newcommand{\tech}{{\textsc{AutoVerus}}\xspace}
\newcommand{\tool}{{\textsc{AutoVerus}}\xspace}

\newcommand{\parabf}[1]{\noindent\textbf{#1}}

\newcommand{\CodeIn}[1]{{\small \texttt{#1}}}
\newcommand{\Comment}[1]{}

\newcommand{\llm}{{LLM}\xspace} 
\newcommand{\llms}{{LLMs}\xspace}
\newcommand{\gpt}{{GPT-}\xspace} 
\newcommand{\houdini}{{Houdini}\xspace} 

\newcommand{\phaseone}{{Phase-1}\xspace} 
\newcommand{\phasetwo}{{Phase-2}\xspace} 
\newcommand{\phasethree}{{Phase-3}\xspace}

\newcommand{\codedelete}{red!60}

\definecolor{codegreen}{rgb}{0,0.6,0}
\definecolor{codegray}{rgb}{0.5,0.5,0.5}
\definecolor{codepurple}{rgb}{0.58,0,0.82}
\definecolor{backcolour}{rgb}{0.97,0.97,0.95}
\definecolor{forestgreen}{rgb}{0.28,0.62,0.37}
\definecolor{codeblue}{rgb}{0,0.5,1}

\lstdefinestyle{mystyle}{
    backgroundcolor=\color{backcolour},   
    commentstyle=\color{codepurple},
    keywordstyle=\color{codepurple},
    numberstyle=\tiny\color{codegray},
    stringstyle=\color{blue},
    basicstyle=\ttfamily\scriptsize,
    breakatwhitespace=false,         
    breaklines=true,                 
    captionpos=b,                    
    keepspaces=true,                 
    numbers=left,                    
    numbersep=5pt,                  
    showspaces=false,                
    showstringspaces=false,
    showtabs=false,                  
    tabsize=4,
}

\lstdefinestyle{ruststyle}{
    morekeywords={
        as, break, const, continue, crate, else, enum, extern, false, fn, for, if, impl, in, let, loop, match, mod, move, mut, pub, ref, return, self, Self, static, struct, super, trait, true, type, unsafe, use, where, while, dyn, abstract, alignof, become, box, do, final, macro, offsetof, override, priv, proc, pure, sizeof, typeof, unsized, virtual, yield, async, await, try
    },
    sensitive=true, %
    morecomment=[l]{//},  %
    morecomment=[s]{/*}{*/},  %
    morestring=[b]",  %
    morestring=[b]{'}, %
    keywordstyle=\color{codepurple},  %
    commentstyle=\color{codegray}\itshape,  %
    stringstyle=\color{blue},  %
    identifierstyle=\color{black},  %
    ndkeywordstyle=\color{purple}\bfseries,  %
    basicstyle=\ttfamily\scriptsize,  %
    showstringspaces=false,  %
    tabsize=4,  %
    breaklines=true,  %
    breakatwhitespace=false,  %
    showtabs=false,  %
    showspaces=false,  %
    showstringspaces=false,  %
    numbers=left,  %
    numberstyle=\tiny\color{gray},  %
    numbersep=-0.1pt,
    xleftmargin=4pt,
}

\lstdefinelanguage{Verus}{
    style=ruststyle,
    morekeywords=[2]{ requires, ensures, invariant, spec, proof, decreases },
    keywordstyle=[2]\color{red}
}

\newcommand{\VerusBG}{\makebox[0pt][l]{\color{YellowOrange!50}\rule[-0.45em]{\linewidth}{1.3em}}}
\newcommand{\VerusBGlight}{\makebox[0pt][l]{\color{Yellow!50}\rule[-0.45em]{\linewidth}{1.3em}}}
\newcommand{\VerusDelete}{\makebox[0pt][l]{\color{red!20}\rule[-0.45em]{\linewidth}{1.3em}}}
\newcommand{\VerusAdd}{\makebox[0pt][l]{\color{green!20}\rule[-0.45em]{\linewidth}{1.3em}}}

\lstdefinelanguage{Markdown}{
    morekeywords={-, *, **},
    sensitive=true,
    morecomment=[l]{<!--}, %
    morecomment=[s]{```}{```}, %
    morestring=[b], %
}

\lstdefinestyle{markdownStyle}{
    language=Markdown,
    basicstyle=\ttfamily,
    keywordstyle=\color{blue},     %
    commentstyle=\color{codegray},     %
    stringstyle=\color{red},       %
}

\AtBeginDocument{%
  }

\begin{document}

\title{\tech: Automated Proof Generation for Rust Code}

\author{Chenyuan Yang}
\email{cy54@illinois.edu}
\orcid{0000-0002-7976-5086}
\affiliation{%
  \institution{University of Illinois at Urbana-Champaign}
  \country{USA}
}

\author{Xuheng Li}
\orcid{0009-0000-1371-2179}
\email{xuheng@cs.columbia.edu}
\affiliation{%
  \institution{Columbia University}
  \country{USA}
}

\author{Md Rakib Hossain Misu}
\orcid{0000-0002-7931-6782}
\email{mdrh@uci.edu}
\affiliation{%
  \institution{University of California Irvine}
  \country{USA}
}

\author{Jianan Yao}
\orcid{0009-0008-4675-8980}
\email{jy3022@columbia.edu}
\affiliation{%
  \institution{University of Toronto}
  \country{Canada}
}

\author{Weidong Cui}
\email{weidong.cui@microsoft.com}
\orcid{0000-0002-2871-9485}
\affiliation{%
  \institution{Microsoft Research}
  \country{USA}
}

\author{Yeyun Gong}
\email{yegong@microsoft.com}
\orcid{0000-0001-9954-9674}
\affiliation{%
  \institution{Microsoft Research Asia}
  \country{China}
}

\author{Chris Hawblitzel}
\orcid{0000-0002-5676-0362}
\email{Chris.Hawblitzel@microsoft.com}
\affiliation{%
  \institution{Microsoft Research}
  \country{USA}
}

\author{Shuvendu Lahiri}
\orcid{0000-0002-4446-4777}
\email{shuvendu@microsoft.com}
\affiliation{%
  \institution{Microsoft Research}
  \country{USA}
}

\author{Jacob R. Lorch}
\email{jaylorch@gmail.com}
\orcid{0000-0002-7269-2769}
\affiliation{%
  \institution{Microsoft Research}
  \country{USA}
}

\author{Shuai Lu}
\email{shuailu@microsoft.com}
\orcid{0000-0001-7466-2064}
\affiliation{%
  \institution{Microsoft Research Asia}
  \country{China}
}

\author{Fan Yang}
\email{fanyang@microsoft.com}
\orcid{0000-0002-0378-060X}
\affiliation{%
  \institution{Microsoft Research Asia}
  \country{China}
}

\author{Ziqiao Zhou}
\email{ziqiaozhou@microsoft.com}
\orcid{0009-0007-2762-0989}
\affiliation{%
  \institution{Microsoft Research}
  \country{USA}
}

\author{Shan Lu}
\email{shanlu@microsoft.com}
\orcid{0000-0002-0757-4600}
\affiliation{%
  \institution{Microsoft Research}
  \city{Redmond}
  \country{USA}
}
\affiliation{%
  \institution{University of Chicago}
  \city{Chicago}
  \country{USA}
}



%%% The following is specific to OOPSLA2 '25 and the paper
%%% 'AutoVerus: Automated Proof Generation for Rust Code'
%%% by Chenyuan Yang, Xuheng Li, Md Rakib Hossain Misu, Jianan Yao, Weidong Cui, Yeyun Gong, Chris Hawblitzel, Shuvendu K. Lahiri, Jacob R. Lorch, Shuai Lu, Fan Yang, Ziqiao Zhou, and Shan Lu.
%%%
\setcopyright{cc}
\setcctype{by}
\acmDOI{10.1145/3763174}
\acmYear{2025}
\acmJournal{PACMPL}
\acmVolume{9}
\acmNumber{OOPSLA2}
\acmArticle{396}
\acmMonth{10}
\received{2025-03-26}
\received[accepted]{2025-08-12}

\renewcommand{\shortauthors}{Yang et al.}

\begin{abstract}
Generative AI has shown its value for many software engineering
tasks. 
Still in its infancy, large language model (\llm)-based proof generation lags behind \llm-based
code generation.
In this paper, we present \tool. \tool uses \revision{\llm{s}} to automatically generate
correctness proof for 
Rust code. 
\tool is designed to match the unique features of Verus, a verification tool 
that can prove the correctness of Rust code using proofs and specifications also
written in Rust. \tool consists of
a network of agents that are crafted and orchestrated to mimic
human experts' three phases of proof construction: preliminary
proof generation, proof refinement guided by generic tips, and proof debugging
guided by verification errors. To thoroughly evaluate \tool and help foster future research in this direction, we have built a benchmark suite of 150 non-trivial proof tasks, based
on existing code-generation benchmarks and verification benchmarks. Our evaluation
shows that \tool can automatically generate correct proof for more than 90\% of them, with more than
half of them tackled in less than 30 seconds or 3 \llm calls.

\end{abstract}

\maketitle

\section{Introduction}
Generative AI (GenAI) techniques have been shown to be effective for many software engineering tasks. GenAI-based code-synthesis tools, like GitHub Copilot, have been widely used in practice. A recent survey among professional developers has shown that more than 60\% of developers are currently using AI-assistants in their code development, and yet two thirds of them do not trust the code generated by AI \cite{stackoverflow}. It would be great if GenAI could synthesize not only code but also code-correctness proof. Indeed, recent
research \cite{DBLP:conf/nips/YangSGCSYGPA23,DBLP:journals/corr/abs-2405-01787,DBLP:conf/saiv/SunSPB14,DBLP:journals/pacmse/MisuLM024, DBLP:journals/corr/abs-2405-16792,DBLP:journals/corr/abs-2406-08467,DBLP:journals/corr/abs-2404-10362} has shown the potential of GenAI in synthesizing proofs in proof-oriented languages like LEAN \cite{DBLP:conf/cade/Moura021}, F$^*$ \cite{DBLP:conf/popl/SwamyHKRDFBFSKZ16}, Dafny \cite{DBLP:conf/lpar/Leino10}.
However, the state of the art of GenAI-for-proof is still far behind that of GenAI-for-coding both in terms of benchmark building and generation capability. Furthermore, there has been little work about how GenAI can help directly prove the correctness of programs written in a general-purpose popular programming language. 
In this paper, we explore using \revision{\llm{s}} to 
automatically generate proof annotations that allow Verus \cite{DBLP:journals/pacmpl/LattuadaHCBSZHPH23} to formally verify the correctness of Rust code.

We choose to focus on Verus for two main reasons.
First, the usage opportunity: Verus is a state-of-the-art SMT-based verifier for Rust with a particular focus on practical usage
\cite{DBLP:conf/osdi/0013MGMCH0PSSX24, DBLP:conf/osdi/ZhouACGHC24, verussosp24}. Verus directly works on Rust code;
Verus specifications and proof annotations are also written with Rust syntax and are inlined in the Rust code under proof in the form of \textit{ghost code}. 
This way, users do not need to learn a new language. Verus 
also leverages the Rust type system to avoid the complexity of reasoning memory-safety and aliasing,
largely reducing the amount of needed proof annotation. 
Considering the growing popularity of Rust \cite{ONCD2024, mitreviewworldfastestgrowingpl,stackoverflowRust}, an effective proof assistant for Verus has the potential to benefit millions of developers.

Second, the research opportunity: Verus has unique features that present intriguing research questions for GenAI. Unlike many verifiers, Verus
does not require \revision{\llms} to learn a new language --- \llms should be familiar with Rust already. However, also unlike many established verifiers, Verus is very young, with its development started about 3 years ago, and hence
has much fewer examples for \revision{\llms} to learn from. 
Furthermore, although Verus proof annotations (short as proof) are written in Rust syntax, their usage is
very different from regular Rust executable code and require 
expertise in verification to master.
Verus' emphasis on verification speed, a property that is critical for Verus to work for large system code, sometimes comes at the cost of extra proof annotations that may not be needed by other SMT-based verifiers (e.g., Dafny), which creates extra challenges for GenAI. 

\begin{figure}

\begin{minipage}[t]{0.4\linewidth}
\centering
\begin{subfigure}[t]{\linewidth}
\begin{lstlisting}[language=Verus, escapechar=!]
!\VerusBGlight!spec fn is_digit(c: u8) -> bool {
!\VerusBGlight!    c >= 48 && c <= 57
!\VerusBGlight!}

!\VerusBGlight!spec fn cnt_dig(seq: Seq<u8>) -> int
!\VerusBGlight!    decreases seq.len(),
!\VerusBGlight!{
!\VerusBGlight!    if seq.len() == 0 {
!\VerusBGlight!        0
!\VerusBGlight!    } else {
!\VerusBGlight!        cnt_dig(seq.drop_last()) + 
!\VerusBGlight!        if is_digit(seq.last()) {
!\VerusBGlight!            1 as int
!\VerusBGlight!        } else {
!\VerusBGlight!            0 as int
!\VerusBGlight!        }
!\VerusBGlight!    }
!\VerusBGlight!}
\end{lstlisting}
    \caption{The spec functions.}
    \label{fig:spec_func}
\end{subfigure}
\end{minipage}
\hfill
\begin{minipage}[t]{0.57\linewidth}
\centering
\begin{subfigure}[t]{\linewidth}
\begin{lstlisting}[language=Verus, escapechar=!, , firstnumber=19]
fn count_digits(text: &Vec<u8>) -> (ret: usize)
!\VerusBGlight\label{line:sum-post-cond}!    ensures  ret == cnt_dig(text@),
{
    let mut count = 0;
    let mut i = 0;
    while i < text.len()
!\VerusBG!      invariant
!\VerusBG\label{line:sum-loop-start}!        i <= text.len(),
!\VerusBG\label{line:sum-loop-mid}!        count <= i,
!\VerusBG\label{line:sum-loop-end}!        count==cnt_dig(text@.subrange(0, i as int)),
    {
      if text[i] >= 48 && text[i] <= 57
         {count += 1;}
      i += 1;
!\VerusBG\label{line:assert1}!      assert(text@.subrange(0, i - 1 as int) 
!\VerusBG!        == text@.subrange(0, i as int).drop_last());
    }
!\VerusBG\label{line:assert2}!    assert(text@ == text@.subrange(0, i as int));
    count
}\end{lstlisting}
\vspace{-0.1in}
    \caption{The implementation.}
    \label{fig:impl_func}
\end{subfigure}
\end{minipage}

    \caption{A Rust function, in \colorbox{backcolour}{gray} background, with Verus annotations. The \colorbox{YellowOrange!50}{dark-yellow} background highlights the proof annotation needed by Verus to prove the specifications highlighted in the \colorbox{Yellow!50}{light-yellow}.}
    \label{fig:sum}

\end{figure}

\subsection{An Example}

Figure \ref{fig:sum} illustrates an example of Rust code annotated with 
Verus specification and proof, which we will refer to
as \textit{Verus program} in this paper. This particular example is based on Task 764 in the
MBPP dataset \cite{DBLP:journals/corr/abs-2108-07732}, which says ``Write a function to count the number of digits in a given string''.
This is exactly what the Rust function \CodeIn{count\_digits} in
Figure~\ref{fig:impl_func} does it through a while loop. 

This function's correctness specification includes two parts: 1)
a spec-function \CodeIn{cnt\_dig} implementing the same
functionality in a recursive manner in Figure~\ref{fig:spec_func}; 2) 
a post-condition of \CodeIn{count\_digits} (Line \ref{line:sum-post-cond}) stating that its result should be the same
as that of spec-function \CodeIn{cnt\_dig}. 

For Verus to successfully verify this post-condition, two types of proof annotations are needed: 
1) a number of loop invariants
that state what properties are true at the beginning and
the end of each loop iteration (Line~\ref{line:sum-loop-start}--\ref{line:sum-loop-end}); 2) 
two \CodeIn{assert} statements that help the underlying SMT-solver to reason about
the collection \CodeIn{text} (Line \ref{line:assert1}--\ref{line:assert2}). Note that, these
\CodeIn{assert}s are different 
from the Rust builtin \CodeIn{assert!} macro. The Rust \CodeIn{assert!} is used for run-time checking. Instead, the expression enclosed in Verus \CodeIn{assert} is to be verified statically and to provide
hints to the SMT solver.

We use this example to highlight a few points mentioned earlier, and we will offer more explanation about this example in the next section. 

First, the goal of this work is to automatically generate proof annotations when provided with Rust code and its specifications. For this example, we would aim to generate all the loop invariants and the \CodeIn{assert} statements (i.e., all the content in \colorbox{YellowOrange!50}{dark-yellow} background in Figure~\ref{fig:impl_func}). 

Second, luckily, we do not need to teach \llms a new language. As shown in Figure \ref{fig:sum}, except for a few keywords like \CodeIn{ensures},
\CodeIn{invariant}, \CodeIn{spec}, everything else is written in Rust syntax. 

Third, there are still many challenges to address. In this example, why those two \CodeIn{assert} statements are needed and why the
`@' symbol is needed on Line \ref{line:sum-post-cond}, \ref{line:sum-loop-end}, \ref{line:assert1}--\ref{line:assert2}, can be 
difficult to understand for inexperienced Verus users and \llms. The loop invariants in this example are not as difficult, but still non-trivial. For example, it is not obvious why the invariant 
\CodeIn{count <= i} is needed on Line \ref{line:sum-loop-mid} --- it is needed so that Verus can prove
there will be no arithmetic overflow inside the loop.
Things get more complicated when proof functions and quantifiers
(i.e., \CodeIn{forall}, \CodeIn{exists}) are needed as we will see later.

\subsection{Contributions}
Facing these unique opportunities and challenges, we have designed a tool based on GenAI, called \textbf{\tool} \includegraphics[scale=0.062]{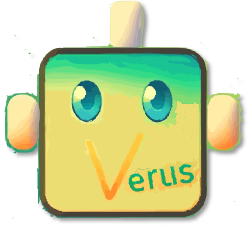}, with the
following principles in mind:

\textit{No reliance on large training data sets}. Unlike prior GenAI-for-proof work that
trains a fine-tuned model \cite{DBLP:conf/sigsoft/FirstRRB23,DBLP:conf/popl/SwamyHKRDFBFSKZ16} or builds a Retrieval Augmented Generation system to dynamically pick different
examples for different proof tasks \cite{DBLP:conf/nips/YangSGCSYGPA23,DBLP:journals/pacmse/MisuLM024}, 
we rely on the code understanding and code synthesis capability of existing \llms (e.g., \gpt{4o}).

\textit{Using human \textbf{Expertise} \includegraphics[scale=0.3]{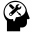} to compensate for the lack of training data}. Specifically,
we summarize common proof strategies used
by Verus experts and put them into the instructions of a set of
\llm agents. We also design \tool to orchestrate  these agents following the proof-development process of 
Verus experts.
Human experts do \textit{not} aim to write a perfect
proof like the one in Figure \ref{fig:sum} in one attempt, neither does \tool. 

\textit{Unleashing the \textbf{Creativity} \includegraphics[scale=0.3]{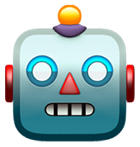} of \llms when expert knowledge does not help}. 
It is impossible to enumerate all proof-writing and debugging tricks, or to anticipate all mistakes that \llm may make. To handle 
inevitable and
    yet unexpected challenges in writing a proof, \tool configures all its agents with a high-temperature,
multi-output setting so that a wide range of diverse and creative proofs can be generated for \tool to pick from. 

\textit{Enforcing \textbf{Discipline} \includegraphics[scale=0.3]{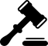} upon creative \llms through formal methods and static checking}. A network of creative \llm agents can produce a huge number of proof candidates\footnote{Imagine a set of $10$ \llm agents each producing $5$ \verus programs based on an input \verus program. Depending on which agent's output to feed into which other agent, there could be more than $10! \times 6^{10}$ proof candidates to choose from!}. Many, if not most, of them are too flawed to further explore and may contain ``cheating behaviors'' like modifying specifications, modifying the Rust code under proof, adding \CodeIn{assume(...)}\footnote{\texttt{assume(P)} creates an assumption that $P$ is true. It is used for proof debugging and should not appear in the final proof.}, etc. To effectively navigate through this huge search space, we combine formal methods and static analysis to quickly filter out invalid candidates, to rank proof candidates based on \verus' feedback, and to stitch proof snippets together in the hope of creating a perfect proof out of a collection of imperfect ones.

\begin{figure}
    \centering
    \includegraphics[width=0.95\linewidth]{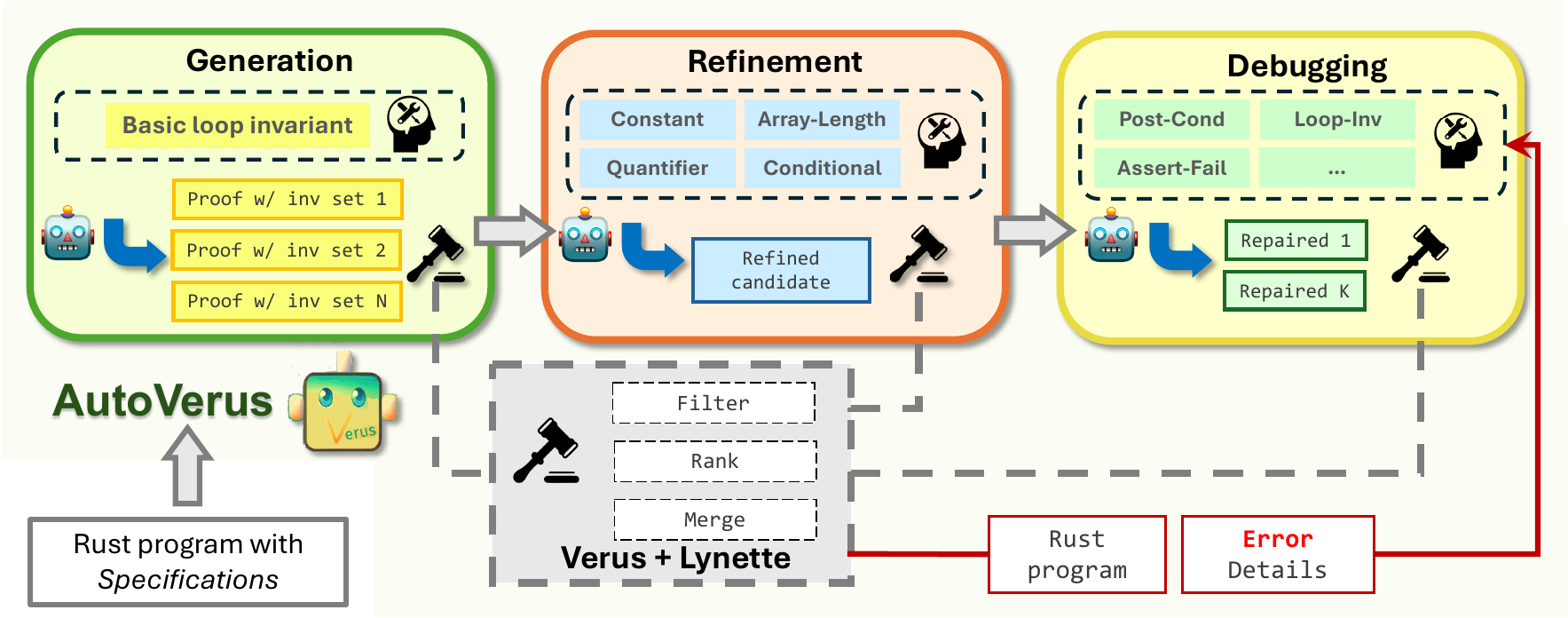}
    \caption{The workflow of \tech{}}
    \label{fig:workflow}
\end{figure}

Putting these together, \tool conducts its proof generation for input Rust code\footnote{In our current prototype,
a single Rust function is the unit of proof generation.} 
in three phases, 
as shown in Figure \ref{fig:workflow}:

\begin{enumerate}
    \item \textit{Generation phase}: \tool uses an \llm agent to generate preliminary loop invariants;
    \item \textit{Refinement phase}: \tool runs a series of agents, each looking for a common mistake or a common omission in loop invariants and conducting refinement accordingly;
    \item \textit{Debugging phase}: Guided by Verus, \tool goes through rounds of debugging to produce the correct proof, not only fixing loop invariants but also adding proof annotations like \CodeIn{assert} statements and lemma functions. In each round,
    \tool examines all the verification errors reported by Verus, decides which error to focus on, and invokes a corresponding agent to fix the error---around 10 agents are designed, each handling one major type of verification errors 
following the common strategies used by Verus experts
(\eg ``post-condition not satisfied'',
``loop invariant not satisfied before the loop'', etc.).

\end{enumerate}

Throughout these three phases, proof filtering, ranking, and merging are
conducted using Verus and Lynette, an AST-level code analysis
tool that we build for \tool upon Verus parser.

To thoroughly evaluate \tool and also to facilitate future research in AI-for-Verus,
we have curated a benchmark suite that consists of 150 non-trivial proof tasks, with each task being a small Rust program associated with
Verus specification.
This is the first benchmark suite designed for Verus proof generation, and we refer to it as \textbf{\verusbench}.
These 150 proof tasks are mainly translated from other benchmark suites
(Diffy \cite{DBLP:conf/cav/ChakrabortyGU20}, MBPP \cite{DBLP:journals/pacmse/MisuLM024}, and CloverBench \cite{DBLP:conf/saiv/SunSPB14})
into Rust/Verus.
More details of \verusbench are in \S~\ref{sec:meth}.

Our evaluation on \verusbench shows encouraging results:
\begin{itemize}
\item \tool has a good proof-synthesis capability. It automatically proves more
than 90\% of the tasks in \verusbench (137 out of 150).
In contrast, a baseline design of directly prompting
\gpt{4o} can only prove 45\% of these tasks (67 out of 150), although this baseline has a much longer time budget, a larger \llm-call budget, a carefully designed prompt, and an unfair advantage of including the answer proof to \revision{four} complicated \verusbench
tasks as examples. 

\item \tool is efficient in its proof synthesis. Even with
a tight budget of 30 seconds or 3 \llm calls
for each task, \tool can figure out the correct proof for more than half of
the \verusbench tasks, while the baseline can
manage fewer than 40 tasks.

\item \tool is robust and consistent across proof tasks and models. It not only successfully finishes 100\% of the \verusbench tasks coming from
Diffy and CloverBench, but also 87\% of the more complicated tasks coming from MBPP, which often require proof annotations that go beyond loop invariants. The \tool framework, including
its three-phase proof generation methodology and proof filtering, ranking, merging support,
works well with different LLM models: GPT-4o, GPT-4-turbo, and Deepseek-R1 all achieve
similar results, with DeepSeek-R1 proving slightly fewer tasks than GPT-4o/turbo, and GPT-4-turbo spending more time in proving the same number of tasks than GPT-4o.

\end{itemize}

Of course, \tool is still just a starting point in our pursuit of GenAI-for-proof. We hope our \verusbench will help evaluate and guide future research in this domain. We also hope our experience of combining human expertise, LLMs, and compiler and formal methods in \tool will be useful for future work in GenAI-for-proof.

\label{sec:intro}

\section{Background: Verus}
Verus verification incurs \textit{no} run-time checks. Instead, Verus statically
analyzes the Rust executable code and the Verus annotations, which are referred to
as \textit{ghost code}, and constructs
SMT formulas to query the underlying SMT solver. Based on the query result, Verus knows if the Rust code can satisfy some specifications for all possible executions. After the verification, Verus can erase all the ghost code to reproduce the original Rust implementation,
allowing easy integration with other Rust code that developers may choose not to verify. This workflow is
unlike many other verification tools that require the verification to be discharged in a different project as the implementation.

Verus has made small extensions on Rust syntax through macros, including new abstract data types. 
Since fixed-bit-width integer types in Rust (e.g., \CodeIn{u64}, 64-bit unsigned integer)
are not easy for SMT solvers to reason about, Verus additionally supports \CodeIn{nat} for any natural numbers and \CodeIn{int} for any integer
numbers. By default, \CodeIn{int} data type is used in all Verus specifications.
Similarly, Verus
provides some collection types (\CodeIn{Seq}, \CodeIn{Map}, \CodeIn{Set}) that can be used to 
abstract Rust collections.

Generally speaking, Verus specification annotations include function pre-condition (\CodeIn{requires}), 
function post-condition (\CodeIn{ensures} as in Figure \ref{fig:impl_func}),
and spec functions. The spec function in Figure \ref{fig:spec_func} looks
 very similar to native Rust functions. However, being ghost code, it is
required to be purely functional without mutations of any variables. This feature allows it to be easily transformed into a function in the SMT solver and can be called from other ghost code, such as in the function post-condition on Line~\ref{line:sum-post-cond},
and in the loop invariant on Line~\ref{line:sum-loop-end}. Note that, executable functions can\textit{not} be
called in ghost code --- a mistake that \llm makes a lot.

Verus proof annotations generally include loop invariants, \CodeIn{assert} statements, and proof functions.
Loop invariants are needed for the verifier to reason about a loop. \CodeIn{assert}s are generally
used to help SMT solvers reason about complicated properties involving quantifiers, collections, etc.
For example, 
a lot of commonly used \verus data-structures (e.g., \CodeIn{Seq}, \CodeIn{Set}) and their specification APIs (e.g., \CodeIn{subrange}, \CodeIn{take}, \CodeIn{filter}, \CodeIn{push}) require accompanying axioms for
SMT solvers to reason about. To avoid complexity explosion in SMT solver, many of these axioms are not 
     automatically triggered in Verus and need to be explicitly added through \CodeIn{assert}s, like that
     in Figure \ref{fig:impl_func}. 

Just like spec functions, proof functions are ghost code and need to be
purely functional. Proof functions
can come with its own premises, specified using \CodeIn{requires}, and conclusions, specified using \CodeIn{ensures}. The body of the proof function contains ghost code used to assist Verus in constructing SMT formulas to prove the conclusions.
A proof function can be called by other ghost code. As long as the premise of the function is satisfied
beforehand, the conclusion of the proof function can help 
prove properties after its call site. We will see an example of proof function later in \S~\ref{sec:phase3}.

\label{sec:back}

\section{Unique Challenges and Opportunities for \tool}

\parabf{Training data and benchmarks.} 
At the time of writing, there are fewer than 10 projects on GitHub developed under Verus. In comparison, there are hundreds of GitHub projects with thousands to hundreds of thousands of files developed under \textit{each} of those more established verification tools like LEAN \cite{DBLP:conf/cade/Moura021}, Coq \cite{coq}, Dafny \cite{DBLP:conf/lpar/Leino10}, 
Isabelle \cite{isabelle}, and F$^*$ \cite{DBLP:conf/popl/SwamyHKRDFBFSKZ16}.
Making things worse, Verus has been designed to verify large system
projects (e.g., storage system
\cite{multiloggit}, VM hypervisor component \cite{DBLP:conf/osdi/ZhouACGHC24},
operating system modules \cite{DBLP:conf/kisv/Chen0MNB23}). Even with our best efforts,
extracting single-function/file benchmark
programs from these projects failed due to complicated
code dependency and specifications.

\parabf{Language syntax.} Unlike most verification tools, Verus allows
\llms to work with a widely used language, Rust. However, this also 
produces challenges, as LLMs easily stumble at those subtle
syntax extensions made by Verus upon Rust.

One such challenge is about the integer data type. Verus uses its abstract \CodeIn{int} type by default
in specification functions and hence
often requires type casting, such as the `\CodeIn{i as int}' on Line \ref{line:sum-loop-end}, 
\ref{line:assert1}--\ref{line:assert2} in Figure \ref{fig:sum}. 
This type of casting is difficult for LLMs like GPT that are used to Rust syntax. 

As another example, since Rust executable functions cannot be called in ghost code, misuse of
Rust APIs in Verus annotation is very common for inexperienced users
and
\llms. In lucky cases, like Line \ref{line:sum-loop-start} of Figure \ref{fig:sum}, 
Verus provides a spec-function \CodeIn{len()} for the Rust collection type \CodeIn{Vec}.
Therefore, \CodeIn{text.len()} can be used as both a valid Rust API call and a valid
spec call in ghost code. Unfortunately, many other Rust APIs do not have corresponding spec.
For example, there is no \CodeIn{subrange} spec function for Rust \CodeIn{Vec} type. 
In Figure \ref{fig:impl_func}, if we use \CodeIn{text.subrange} inside the loop invariants
or the \CodeIn{assert} statement, compilation errors will rise. Instead, \CodeIn{text@.subrange}
is used, where \CodeIn{text@} turns \CodeIn{text} from Rust \CodeIn{Vec} into Verus
\CodeIn{Seq}, which has a \CodeIn{subrange} spec function. There are also cases where
even \CodeIn{@} does not help.
For instance, \gpt{4o} often tries to create a Verus sequence using \CodeIn{!seq[...]} or access the first \CodeIn{x} elements with \CodeIn{seq[..x]}, both of which follow Rust syntax but are invalid for Verus \CodeIn{Seq} data type.

\parabf{What constitutes a proof.} For different types of verification tools, what users need to write to accomplish a
proof can be vastly different, which creates different proof-automation challenges.

Using an interactive theorem prover, such as Coq \cite{coq} and LEAN
\cite{DBLP:conf/cade/Moura021}, the user writes proof tactics to move the verification forward step by step. In contrast, the SMT solver does the heavy-lifting in Verus and the user only provides hints to help the SMT query. In most cases, this means less effort from Verus users. However, this also implies challenges in measuring progress.
When using an interactive theorem prover, the verification goal is transformed based on the tactics used and it is easier to evaluate whether each tactic moves the verification forward, backward, or stagnant by comparing the verification goal before and after the tactic is applied; in contrast, although Verus provides the information regarding which part of the code has caused the constructed SMT formula to fail to hold, it is difficult to precisely tell whether progress has been made towards the end goal of proving function post conditions, a challenge shared by other non-interactive verifier \cite{DBLP:journals/corr/abs-2405-01787}.

Using a proof-oriented language like F$^*$, code implementation and proof are intertwined; in contrast, in Verus, code implementation 
and proof are separated. On one hand, this means an easier task for
\tool, as it only needs to generate the proof but not the implementation unlike that in GenAI-for-F$^*$ \cite{DBLP:journals/corr/abs-2405-01787}. 
On the other hand, this
raises the challenge of disciplining \revision{\llms} into not modifying the given implementation. This turns out to be difficult. For example, when given the
code in Figure \ref{fig:changecode_before}, GPT-4 tends to change it to
be that in Figure \ref{fig:changecode_after}.

\begin{figure}

    \begin{minipage}[t]{0.47\linewidth}
\centering
\begin{lstlisting}[language=Verus]
	while i < N {
		if i == 0 {
			sum.set(0, 0);
		} else {
			sum.set(0, sum[0] + a[i]);
		}
		i = i + 1;
	}
\end{lstlisting}
\vspace{-0.1in}
    \caption{A code snippet from Diffy benchmark}
   \label{fig:changecode_before}
\end{minipage}
\hfill
\begin{minipage}[t]{0.47\linewidth}
\centering
    \begin{lstlisting}[language=Verus]
    sum[0] = 0;
    
	while i < N {
		sum.set(0, sum[0] + a[i]);
		i = i + 1;
	}
    \end{lstlisting}
    \caption{GPT-4 output when fed w/ the code in Fig. \ref{fig:changecode_before}}
   \label{fig:changecode_after}
\end{minipage}

\end{figure}

\parabf{Miscellaneous features of Verus.}
The design of Verus has paid particular attention to the verification speed. For example, Verus often takes only a second to verify a function. 
Although this allows \tool to feed many proof candidates to Verus to see which one works, it sometimes comes at the cost of extra annotations of loop invariants and \CodeIn{assert}s.

\label{sec:cha}

\section{Design}
\textit{Terminology.} The input and output of \tech, as well as each of its agents, are \textit{Verus programs}
that contain Rust code and Verus annotations.
The input program to \tech contains no proof annotations; the output of \tech contains not only 
Rust code, specification, but also proof annotations added by it.
We refer to a Verus program as \textit{unsafe}, if it modifies the Rust code and/or 
specifications under proof; as \textit{invalid}, if it leads to Verus compilation errors; as \textit{correct}, if it is
safe, valid, and can be completely verified by Verus.

\revision{To design \tool, we have interviewed multiple co-authors of the Verus paper \cite{verussosp24}. Their suggestion about how to handle common Verus errors guided the design of \tool agents. Furthermore, their description of the
common proof-development process laid the foundation of the three-phase workflow. Specifically, we learned that the proof development of human experts is an iterative process of repeatedly running \verus, checking and prioritizing \verus errors, developing and editing proof to fix them, and so on---a process that is also confirmed through
user studies \cite{TODO:userstudy}.}

In the following, we present each of the three iterative phases of AutoVerus, discussing how the
LLM agents of each phase are designed, how they are coordinated, and how their output is processed leveraging Verus.

\begin{algorithm}[t]

\small
\caption{Preliminary proof generation with post-processing}
\label{alg:prelim-proof}
\DontPrintSemicolon
\SetKwProg{Fn}{Function}{:}{}
\SetKwFunction{GenPrelimProof}{GenPrelimProof}
\SetKwFunction{LoopInvGen}{LoopInvAgent}
\SetKwFunction{IsSafe}{is\_safe}
\SetKwFunction{Score}{score}
\SetKwFunction{Merge}{merge}
\SetKwFunction{IsCorrect}{is\_correct}
\SetKwFunction{Houdini}{houdini}
\SetKwFunction{VerusCorrect}{verus\_correct}

\Fn{\GenPrelimProof{program}}{
    $pCandidates \gets$ \LoopInvGen{program}\;
    $safeCandidates \gets \{\, p \in pCandidates \mid \IsSafe(program, p)\}$\; \label{algo:phase1-filter}
    {Sort $safeCandidates$ by \Score}\; \label{algo:phase1-sort}
    $mergeP \gets program$\;
    \ForEach{$p \in safeCandidates$}{ \label{algo:phase1-merge}
        $mergeP \gets$ \Merge($mergeP, p$)\; 
        \If{\IsCorrect($mergeP$)}{
            \Return $mergeP$\; \label{algo:phase1-correct}
        }
    }
    \Return first element of $safeCandidates$\; \label{algo:phase1-ret}
}
\Fn{\IsCorrect{$p$}}{
    \Return \VerusCorrect($p$) \textbf{ or } \VerusCorrect(\Houdini($p$))\;
}
\end{algorithm}

\subsection{Phase 1: Preliminary Proof Generation}
\label{sec:phase1}

\verus proof typically includes loop invariants,
\CodeIn{assert}s, and sometimes proof functions/blocks. Since \CodeIn{assert}s and proof functions/blocks are typically added during the proof debugging to provide extra hints to SMT solver, at this first phase of
proof generation, \tool focuses on loop invariants. An agent is designed to generate loop invariants for every
loop in the target Rust code.

\subsubsection{Agent design}
In the prompt to \revision{\llms}, we ask it to
``add loop invariants to the given Rust code, so that \verus can verify the given function behaves exactly what is described in the specifications.'' Furthermore, we teach \revision{\llms} three
things that we believe are the most important in writing loop invariants in \verus: 1) to describe the initial value of every variable read in the loop; 2) to describe the assigned value of every variable written in the loop; and
3) to leverage spec and proof functions when needed. 
We have intentionally kept this agent simple and leave it to later refinement/repair agents to correct any mistakes or
oversights of this first agent.

We also include three toy \verus examples
written by us in the prompt to teach \revision{\llms} about basic \verus features like \CodeIn{invariant},
quantifiers (i.e., \CodeIn{forall} and \CodeIn{exists}), and abstract data structures (e.g., \CodeIn{vstd::Seq}).
Each example contains less than 30 lines of Rust code and 10--20 lines of proof annotations. We stick to these same
examples across all proof tasks.

\subsubsection{Post-Processing of Agent Output} \phaseone of \tool is much more than just invoking an LLM agent, as much post-processing is needed to filter, rank, and merge the LLM agent's output.

As shown in Algorithm~\ref{alg:prelim-proof}, after invoking the LLM agent described above to produce $5$ candidates,
\tool first filters out the candidates that are unsafe, like the one in Figure \ref{fig:changecode_after}, through static analysis.
\tool also filters out invalid candidates that contain syntax errors, unless \tool can fix the errors through simple code editing, such as changing \CodeIn{v[i]} to \CodeIn{v[i as int]} when \verus complains that ``type int is expected for the expression \CodeIn{i}''.

\tool then ranks remaining LLM-generated \verus programs based on a score tuple \{$V$, $E$\}{\xspace (Line~\ref{algo:phase1-sort} in Algorithm \ref{alg:prelim-proof})}, with $V$ being the number of functions\footnote{Verus counts the loop as a separate function. For example, when a function containing two loops is verified, $V$ is $3$.} successfully verified by Verus, and $E$ being the number of verification errors reported by Verus. 
For any two programs that contain the same executable Rust code, \tool considers the one with higher $V$ score as the better one (i.e., more functions proved), with the tie-breaker
being having a lower $E$ score (i.e., fewer verification errors).
If the top-ranked \verus program allows verification to succeed, the whole proof task is done. Otherwise, \tool moves on.

Finally, given a list of $K$ valid and safe programs, $P_1$, $P_2$, ..., $P_k$, \tool checks if merging some of them might produce a better proof {\xspace (Line~\ref{algo:phase1-merge} -- \ref{algo:phase1-correct} in Algorithm \ref{alg:prelim-proof})}. Since merging takes time, \tool only explores linear, instead of exponential, number of merging schemes: \tool starts with the highest ranked program, considered as best-so-far $P_B$, and goes down the ranked list. When merging the best-so-far program $P_B$ with a program $P_i$ produces a higher-scored program $P'$, \tool updates $P_B$ to be $P'$ and continues to check the next program on the
list $P_{i+1}$ until all the $k$ programs are checked. If \tool fails to identify a perfect proof
throughout this process, the final best-so-far program $P_B$ will become the input to the next phase of \tool { (Line~\ref{algo:phase1-ret}).

\begin{figure}

\begin{minipage}[t]{0.47\linewidth}
\centering
\begin{subfigure}[t]{\linewidth}
\begin{lstlisting}[language=Verus, escapechar=!]
invariant
  N > 0,
  i <= N as usize,
  sum.len() == 1,
  sum[0] <= N + i,
  forall|k: int|0<=k<N ==> b[k]==1,
!\VerusBGlight\label{line:merge_a}!  b.len() == N,
// (V, E) = (4, 2)
\end{lstlisting}
\vspace{-0.1in}
    \caption{Loop invariants of best-so-far $P_B$.}
    \label{fig:merge_a}
\end{subfigure}
\begin{subfigure}[t]{\linewidth}
\begin{lstlisting}[language=Verus, escapechar=!]
invariant
  N > 0,
  i <= N as usize,
  sum.len() == 1,
  sum[0] <= N + i,
  forall|k: int|0<=k<N ==> b[k]==1,
!\VerusBG!  N < 1000,!\label{line:merge_b}!
// (V, E) = (3, 2)
\end{lstlisting}
\vspace{-0.1in}
    \caption{Loop invariants of a lower-ranked $P_i$.}
    \label{fig:merge_b}
\end{subfigure}
\end{minipage}
\hfill
\begin{minipage}[t]{0.5\linewidth}
\centering
\begin{subfigure}[t]{\linewidth}
\begin{lstlisting}[language=Verus, escapechar=!]
i = 0;

while (i < N as usize)
  invariant
  N > 0,
  i <= N as usize,
  sum.len() == 1,
  sum[0] <= N + i,
!\label{line:diffy-quant}!  forall|k: int|0<=k<N ==> b[k]==1,
!\VerusBGlight!  b.len() == N,
!\VerusBG!  N < 1000,
{
  sum.set(0, sum[0] + b[i]);
  i = i + 1;
}
// (V, E) = (5, 0)
// Verified!
\end{lstlisting}
    \caption{The merged program $P'$ which is \emph{verified.}}
    \label{fig:merge_c}
\end{subfigure}
\end{minipage}

\caption{An example of a merged program for a problem from \diffy benchmark.}
\label{fig:merge}

\end{figure}

Figure~\ref{fig:merge} shows an example of how \tool handles a problem from \diffy benchmark that
contains 4 loops. 
For this task, 
none of the five programs output by \revision{the \llm} is correct. In fact, two of them
cannot even be compiled, because \llm used quantifiers incorrectly --- it adds
``\CodeIn{forall |k:int| ==> sum[0] == k + i}'' as an invariant for the loop in Figure \ref{fig:merge_c} (there is no way \CodeIn{sum[0]} can equal all integers!). Fortunately, the other three programs are safe and valid. 
In the highest-ranked program, three loops and the function
post-condition are verified (i.e., $V$ score is 4), but two verification errors are reported for the
last loop with the loop invariants shown
in Figure \ref{fig:merge_a}. Another program encountered index out-of-bound errors in the last two loops, and hence is ranked lower (i.e., $V$ is 3). Its invariants for the last loop are shown in Figure \ref{fig:merge_b}.
As we can see, these two sets of invariants only differ by the last line: the lack of 
$N < 1000$ in Figure \ref{fig:merge_a} caused arithmetic overflow errors; the lack of 
$b.len() == N$ in Figure \ref{fig:merge_b} caused its index out-of-bound error. When we
merge the second program (Figure \ref{fig:merge_b}) into the first, we get the perfect proof
in Figure \ref{fig:merge_c}.

Some implementation details are skipped here and will be presented in \S~\ref{sec:impl}: we have implemented a static analysis and code transformation tool, \lynette, to
check whether a \verus program is safe and to merge two \verus programs together
if they are based on the same Rust code. Furthermore,
for every valid and safe \verus program $P$ (no matter generated 
directly by \llm or through merging), we apply \houdini algorithm
\cite{DBLP:conf/fm/FlanaganL01} to efficiently check if a subset of the \verus proof annotations 
in $P$ is a correct proof. 
This \houdini treatment applies to the next two phases as well.

\subsection{Phase 2: Generic Proof Refinement}
\label{sec:phase2}

This phase aims to refine the loop invariants in the best-so-far \verus program $P_B$ generated by phase-1, correcting common verification mistakes and omissions.

\subsubsection{Refinement-Agent Design}

The refinement phase consists of a series of \llm agents, each
aiming one common mistake in the writing of loop invariants in
\verus:

\CodeIn{Constant-Propagation} agent checks if the
    function pre-condition includes properties about a read-only
    parameter, and adds these properties as invariants
    to every loop in the function. 
    For example, loop invariants \CodeIn{N > 0} and \CodeIn{N < 1000} in Figure \ref{fig:merge_c}
    are part of the function pre-condition in that example. If they are missing, this agent will
    add them.

\CodeIn{Array-Length} agent checks if the length information of every array/container used in a loop is specified as a loop invariant, and adds such information if not. For example, if the
invariant \CodeIn{sum.len()==1} is missing from Figure \ref{fig:merge_c}, it is be added here.

\CodeIn{Quantifier} agent checks if quantifier-related invariants are used correctly. 
For example, if Line \ref{line:diffy-quant} of Figure \ref{fig:merge_c} mistakenly has
\CodeIn{forall |k:int| 0 <= k < i ==> b[k] == 1}, instead of \CodeIn{forall |k:int| 0 <= k < N ==> b[k] == 1},
it would be corrected here.

\CodeIn{Conditional-Loop-Invariant} agent urges \llm to check if any loop invariant may only apply to some, instead of all, of the loop iterations and make adjustment accordingly. For instance, this would help \revision{\llms} handle the example in Figure \ref{fig:changecode_before}.

Note that, this set of refinement agents can be easily extended, and the post-processing
presented below makes sure that different agents can easily be composed. In the current prototype of \tool, we keep this set relatively small, as these agents are invoked 
unconditionally regardless of how many and what verification errors are reported by Verus.
These four agents are designed based on common mistakes that we ourselves typically
commit when we write loop invariants.

\subsubsection{Agent Composition and Output Post-Processing}

{Algorithm~\ref{alg:refine-proof} shows the refinement workflow.}
\begin{wrapfigure}{r}{0.5\linewidth}
\begin{minipage}{\linewidth}
\begin{algorithm}[H]
\small
\caption{Proof Refinement}
\label{alg:refine-proof}
\DontPrintSemicolon
\SetKwProg{Fn}{Function}{:}{}
\SetKwFunction{RefineProof}{RefineProof}
\SetKwFunction{IsCorrect}{is\_correct}
\SetKwFunction{AcceptRefine}{accept\_refine}
\SetKwFunction{ConstantRefine}{ConstantRefine}
\SetKwFunction{ArraylenRefine}{ArraylenRefine}
\SetKwFunction{QualifierRefine}{QualifierRefine}
\SetKwFunction{CondloopRefine}{CondloopRefine}
\SetKwFunction{RefineAgent}{refineAgent}
\SetKwFunction{RefineAgentList}{refineAgentList}

\Fn{\RefineProof{program}}{
    $best \gets program$\;
    \ForEach{$\RefineAgent \in \RefineAgentList$}{ \label{algo:phase2-agent}
        $p \gets \RefineAgent(best)$\; \label{algo:phase2-agent-app}
        \If{\IsCorrect($p$)}{
            \Return $p$\; \label{algo:phase2-correct}
        }
        \If{\AcceptRefine($p$)}{\label{algo:phase2-accept}
            $best \gets p$\;
        }
    }
    \Return $best$\;
}
\end{algorithm}
\end{minipage}
\end{wrapfigure}
\tool invokes every refinement agent sequentially {\xspace (Line~\ref{algo:phase2-agent} of Algorithm \ref{alg:refine-proof})}, starting with the simplest, \CodeIn{Constant-Propagation},
and progressing to the more complex ones. Each agent takes the best-so-far \verus program $P_B$ 
as its input and generates a new refined program $P$. If $P$ turns out to contain the correct proof,
the whole task is done; otherwise, \tool replaces the original best-so-far program $P_B$ with
$P$ when two conditions are met:
1) $P$ is a valid and safe \verus program after \tool's type-casting edits, if needed; and, 2) $P$ does not lower the number of functions successfully verified by \verus in $P_B$.

The post-processing here is simpler than that of \phaseone: each agent only generates one, instead of five, proof candidates, and hence there is no merging or ranking needed.
The rationale is that each agent in this phase conducts a relatively straightforward action, and, though every agent is useful in general, it may or may not be necessary for every \verus program given to it.

\begin{figure}[t]
    \begin{minipage}{0.6\linewidth}
        \begin{algorithm}[H]
        \small
        \caption{Debug Proof Generation}
        \label{alg:debug-proof}
        \DontPrintSemicolon
        \SetKwProg{Fn}{Function}{:}{}
        \SetKwFunction{DebugProof}{DebugProof}
        \SetKwFunction{IsCorrect}{is\_correct}
        \SetKwFunction{VerusCompile}{verus\_compile}
        \SetKwFunction{SelectError}{select\_error}
        \SetKwFunction{SelectRepairAgent}{select\_repair\_agent}
        \SetKwFunction{AcceptRepair}{accept\_repair}
        \SetKwFunction{RepairAgent}{repair\_agent}
        \Fn{\DebugProof{program}}{
            \For{$i \gets 1$ \KwTo $MaxIter$}{
                \If{\IsCorrect(program)}{
                    \Return program\;
                }
                $errors \gets$ \VerusCompile(program)\; \label{algo:phase3-all-errors}
                $e \gets$ \SelectError(errors)\; \label{algo:phase3-one-error}
                $repair\_agent \gets$ \SelectRepairAgent($e$)\; \label{algo:phase3-error-agent}
                $repairedCandidates \gets$ repair\_agent(program, $e$)\;
                \ForEach{$p \in repairedCandidates$}{
                    \If{\IsCorrect($p$)}{
                        \Return $p$\;
                    }
                    \If{\AcceptRepair($p$)}{
                        $program \gets p$\; \label{algo:phase3-accept}
                        \textbf{break}\;
                    }
                }
            }
            \Return program\;
        }
        \end{algorithm}
    \end{minipage}
    \hfill
    \begin{minipage}{0.38\linewidth}
        \centering
        \includegraphics[width=0.5\linewidth]{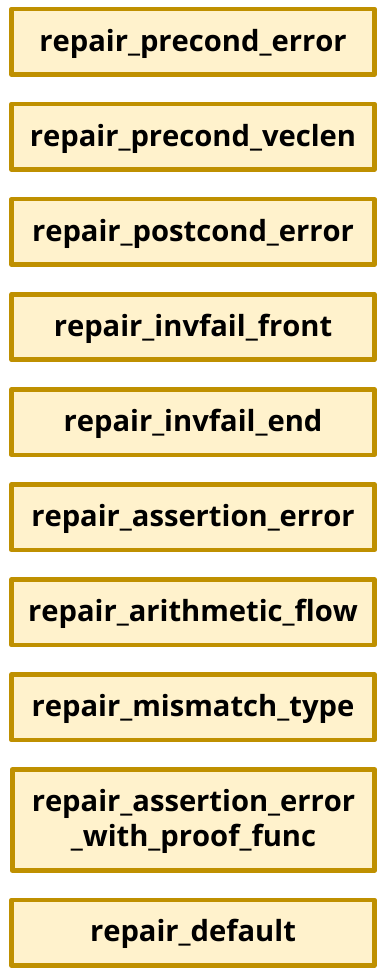}
        \caption{{\tool repair agent list}}
        \label{fig:repair-agents}
    \end{minipage}
\end{figure}
\subsection{Phase 3: Error-Driven Proof Debugging}
\label{sec:phase3}

Proof construction is similar to code writing in that
debugging is inevitable for realistic and complicated tasks. 
Human experts
expect to see verification errors and are experienced
at repairing proofs to fix those errors one by one.
Therefore, we have designed this debugging phase for \tool to
mimic human experts when its first two phases fail to accomplish the proof.

This debugging phase is arguably the most important phase of \tool: any mistake or
limitation of the first two phases can be rescued here. This is also the phase that allows
\tool to generate proof annotations beyond just loop invariants, as we see below.

\subsubsection{Repair-Agent Design} The agent design in this phase is guided by a
list of common Verus verification errors, with each agent $A_t$
designed to fix one type $t$ of \verus verification errors. 

Specifically, \tool targets the eight most common types of verification errors, including function pre-condition errors, vector-index out-of-bound errors (a common sub-type of function pre-condition errors), function post-condition errors, loop-invariant unsatisfied at the beginning of a loop errors, loop-invariant unsatisfied at the end of a loop errors, assertion errors, arithmetic under/overflow errors, and type mismatch (syntax) errors.
As listed in Figure \ref{fig:repair-agents}, one repair agent is designed for each type of error. In addition, an extra repair agent is designed to handle assertion errors, the most diverse type of verification errors. Finally, a generic repair agent, ``repair\_default'' in Figure \ref{fig:repair-agents}, is designed to handle any verification error that does not fall into the eight types listed above.
In practice, it is very rare to encounter verification errors outside these eight types, as we will see in our evaluation.

The prompt to agent $A_t$ that handles a verification error of type $t$ includes not only the \verus
program under repair, but also 1) the code snippet and detailed information about an error $E_t$ to be fixed, which has the verification-error type
$t$, and 2) the instruction about how to fix errors
of type $t$. The instruction is crafted by experienced Verus users.

It could be boring to go through every repair agent one by one, as each agent basically summarizes how human experts handle a type of verification error.
Thus, we skip that discussion here. Later in this section, we will go through
a case study of how \tool works on the proof task \CodeIn{fibonacci}, where we will see what some of the main agents are and how they actually perform in practice.

\subsubsection{Repair-Agent Composition and Output Post-Processing} It is challenging to coordinate all the agents because a Verus program often contains more than one verification error and each error may take multiple attempts to get fixed. Furthermore, fixing one error may introduce another error(s), and it could be difficult to tell whether the repair has made the proof better or worse.

With these challenges in mind, \tool conducts its debugging in iterations, as shown in Algorithm~\ref{alg:debug-proof}, starting with the 
best-so-far Verus program $P_B$ generated by the refinement phase.
In each 
iteration, \tool collects all the verification errors $\mathbf{E}$ in $P_B$ from \verus {\xspace (Line~\ref{algo:phase3-all-errors})},
picks one error $E$ from $\mathbf{E}$ to fix next {\xspace (Line~\ref{algo:phase3-one-error})}, and prompts
a repair agent that suits $E$ {\xspace (Line~\ref{algo:phase3-error-agent})}. \tool then examines the agent's output one by one
to decide whether to accept a repair to replace the best-so-far proof $P_B$ {\xspace (Line~\ref{algo:phase3-accept})}. 
Finally, \tool prepares for its next iteration based on 
whether the target error $E$ has been fixed or not.
\tool keeps running these iterations until a correctly proved program is found or a pre-configured
threshold is reached ($10$ iterations by default). 
We explain these three steps in more detail below.

\parabf{How to pick the error to fix next?}
We have designed an ordered list of \verus error types and
\tool ensures that a verification error is picked only when higher-priority errors
no longer exist for the program under proof. The top one is \textit{type error}, because 
they have to be fixed before \verus can compile the program and conduct verification 
attempt.
The next two are bound-related errors, including the
\textit{index-bound-error} and \textit{arithmetic-overflow/underflow} errors. 
The reason is that once \verus identifies a bound error
for any component of an expression $E$, it will make no further attempt to prove any
property that involves $E$. For example, Verus will not attempt to verify \textit{any}
of the loop invariants if a bound error is found in the loop body.
After that, \textit{invariant-not-satisfied-before-loop} errors get selected, as they are
often easier to fix than other types of errors. Beyond that,
all remaining errors are repaired one by one based on where they show up in the
report of \verus.

\parabf{When to accept a repair?}
Naturally, any repair that introduces unsafe changes or compilation errors is not 
accepted; and, any repair that produces a better verification score than $P_B$ 
can be accepted like that in Phase 1. Beyond that,
\tool would also accept a program $P$ 
as the new best-so-far proof, as long as $P$ has resolved the error-under-repair $E$ without compromising previously verified functions (i.e., the $V$ score should not drop). Since it is difficult to precisely
judge whether $E$ is fixed, \tool counts the number of verification errors of the type of
$E$ in $P$, and considers $E$ to be fixed if that number has dropped from that in
$P_B$.
Once \tool finds an acceptable repair generated by the agent, it does not
check the remaining output of the agent, because it is very difficult, even for human experts, to rank multiple
imperfect repairs.

\parabf{How to adjust for the next repair iteration?}
\tool would review the progress (or lack of) made by the past iteration and adjust
accordingly. If the target error $E$ has not been fixed in this iteration, 
\tool configures the next repair agent to produce 5 programs (3 otherwise).
For assertion-failure errors, the most common type
in later phase of repair, if the target assert error is not fixed after a threshold number
of attempts (3 by default), the more complex assert-repair agent is employed. 
Finally, if the target error fails to be repaired in even more iterations
(5 by default), 
\tool would delete the corresponding proof entity (e.g., a loop invariant or an assert).

\subsubsection{Case Study: \tool on Fibonacci}
\label{sec:fibo}

Figure \ref{fig:repair_fib} contains a Rust function that returns the first $n$
numbers in the fibonacci sequence. It sets the first two numbers in this sequence to be 0 and 1,
and then computes the remaining ones through a while loop (Lines~\ref{line:fib-while-start}--\ref{line:fib-while-end}). The spec function \CodeIn{fibo} in Figure \ref{fig:fibo_spec} provides a strict mathematical specification of the Fibonacci sequence.
What we need to prove for the Rust function 
\CodeIn{fibonacci(n)} is that every number in its result vector matches that produced by the spec
\CodeIn{fibo} under the premise that the $n$th fibonacci number fits in the \CodeIn{i32} data type. As we will see below, it is quite challenging to generate a complete and correct
proof. \tool will accomplish this through several iterations of debugging.

\textit{\textbf{Iteration 1:}} All the gray-background code in Figure \ref{fig:repair_fib}, including the red line \CodeIn{fib.len() == n}, comes from the best-so-far proof $P_B$
from \tool's refinement phase. Verus only reports one error for this version: 
invariant \CodeIn{fib.len() == n} on Line 24 is not satisfied before the loop.

Our agent dedicated to \textit{invariant-not-satisfied-before-loop} is invoked. Its prompt includes these repair strategies: 1) to add an \CodeIn{assert} about this invariant right before the loop; 2) in case of nested loops, to add this invariant to its outer loop when applicable; 3) to modify or delete this loop invariant when it is incorrect or unnecessary. One of this agent's outputs took both the first and the third strategy, as highlighted by 
\colorbox{\codedelete}{{-1}} and \colorbox{green}{{+1}} in Figure \ref{fig:repair_fib}: it deleted this
invariant from Line~\ref{line:fib-loopinv-delete}, and added an \CodeIn{assert(fib.len() == 2}) right before the loop on Line~\ref{line:fib-loopinv-add}, accompanied by a comment explaining its action on Line~\ref{line:fib-loopinv-add-comment}. In fact, deleting this invariant is the right repair action: the length of
\CodeIn{fib} array matches the loop index $i$ in the loop, not always
$n$. Of course, \CodeIn{fib.len() == 2} is correct right before the loop. So, this repair succeeds.

\begin{figure}

\begin{minipage}[t]{0.47\linewidth}
\centering
\begin{subfigure}[t]{\linewidth}
\begin{lstlisting}[language=Verus, escapechar=!]
fn fibonacci(n: usize) -> (ret: Vec<i32>)
  requires
    fibo_fits_i32(n as int),
    n >= 2,
  ensures
    forall|i: int|2 <= i < n ==> 
        #[trigger] ret@[i] == fibo(i),
    ret@.len() == n,
{
  let mut fib = Vec::new();
  fib.push(0);
  fib.push(1);
  let mut i = 2;
!\label{line:fib-loopinv-add-comment}\VerusAdd\colorbox{green}{\textbf{+1}}\textbf{\color{blue}// Assert the invariant right before the loop}!
!\VerusAdd\colorbox{green}{\textbf{+1}}\label{line:fib-loopinv-add}! assert(fib.len() == 2);
!\label{line:fib-while-start}!  while i < n
    invariant
      0 <= 2 <= i <= n,
      fib.len() == i,
      fibo_fits_i32(n as int),
!\label{line:fib-unhold-inv}!      forall|j: int| 0 <= j < i ==> 
          # [trigger] fib[j] == fibo(j),
!\VerusDelete\colorbox{\codedelete}{\textbf{-1}}\label{line:fib-loopinv-delete}!  fib.len() == n;
  {
!\VerusAdd\colorbox{green}{\textbf{+2}}\textbf{\color{blue}// Assert the addition won't overflow an i32}!
!\VerusAdd\colorbox{green}{\textbf{+2}}\label{line:fib-overflow-assert}! assert(fib[i-1]as int + fib[i-2]as int
!\VerusAdd\colorbox{green}{\textbf{+2}}\label{line:fib-overflow-assert-2}!          < 0x8000_0000) by {
!\VerusAdd\colorbox{green}{\textbf{+2}}\label{line:fib-overflow-assert-st}!  assert(fib[i - 1] == fibo(i - 1));
!\VerusAdd\colorbox{green}{\textbf{+2}}!  assert(fib[i - 2] == fibo(i - 2));
!\VerusAdd\colorbox{green}{\textbf{+2}}!  assert(fibo(i-1)+fibo(i-2)==fibo(i));
!\VerusAdd\colorbox{green}{\textbf{+2}}\label{line:fib-overflow-assert-ed}!  lemma_fibo_monotonic(i, n);
!\VerusAdd\colorbox{green}{\textbf{+2}}! };
!\label{line:fib-overflow}!    let next_fib = fib[i - 1] + fib[i - 2];
    fib.push(next_fib);
    i += 1;
!\label{line:fib-while-end}!  }
  fib
}
!\VerusAdd\colorbox{green}{\textbf{+2}}! proof fn lemma_fibo_monotonic(..) {..}
\end{lstlisting}
\caption{Repair the ``the loop invariant not satisfied before loop'' and ``arithmetic overflow'' errors.}
\label{fig:repair_fib}
\end{subfigure}
\end{minipage}
\hfill
\begin{minipage}[t]{0.47\linewidth}
\begin{subfigure}[t]{\linewidth}
\begin{lstlisting}[language=Verus, escapechar=!]
spec fn fibo(n: int) -> nat
    decreases n,
{
    if n <= 0 { 0 }
    else if n == 1 { 1 } 
    else { fibo(n - 2) + fibo(n - 1) }
}
spec fn fibo_fits_i32(n: int) -> bool {
    fibo(n) < 0x8000_0000
}
\end{lstlisting}
    \caption{The spec functions.}
    \label{fig:fibo_spec}
\end{subfigure}
\begin{subfigure}[t]{\linewidth}
\begin{lstlisting}[language=Verus, escapechar=!]
proof fn lemma_fibo_monotonic(n:int, m:int)
  requires n <= m,
  ensures fibo(n) <= fibo(m),
  decreases m - n
{
  if n < m {
!\label{line:fib-lemma}!    lemma_fibo_monotonic(n, m - 1);
!\VerusAdd\colorbox{green}{\textbf{+}}!  assert(fibo(n) <= fibo(m - 1));
!\VerusAdd\colorbox{green}{\textbf{+}}\label{line:fib-lemma-fail}!  assert(fibo(m - 1) <= fibo(m));
  }
}
\end{lstlisting}
    \caption{Repair the post-condition not satisfied error.}
    \label{fig:repair_a}
\end{subfigure}

\begin{subfigure}[t]{\linewidth}
\begin{lstlisting}[language=Verus, escapechar=!]
proof fn lemma_fibo_monotonic(n:int, m:int)
  ...
{
  if n < m {
    lemma_fibo_monotonic(n, m - 1);
    assert(fibo(n) <= fibo(m - 1));
!\VerusDelete\colorbox{\codedelete}{\textbf{-}}!  assert(fibo(m - 1) <= fibo(m));
!\VerusAdd\colorbox{green}{\textbf{+}}!  if m > 1 { !\textbf{\color{blue}// Avoiding the case m == 1 to prevent negative indexing in fibo(m-2)}!
!\VerusAdd\colorbox{green}{\textbf{+}}!    assert(fibo(m-2)+fibo(m-1)==fibo(m));
!\VerusAdd\colorbox{green}{\textbf{+}}!    assert(fibo(m - 1) <= fibo(m));
!\VerusAdd\colorbox{green}{\textbf{+}}!  }
  }
}
\end{lstlisting}
    \caption{Repair the assertion failure error.}
    \label{fig:repair_b}
\end{subfigure}
\end{minipage}
\caption{How \CodeIn{fibonacci} is proved by repair agents (all the \textbf{\color{blue}comments} are output of \gpt{4o} agents as well).}
\label{fig:repair}
\end{figure}

\textit{\textbf{Iteration 2:}}
After the first repair, Verus now reports two new errors: 1) an \textit{arithmetic-overflow} for
\CodeIn{fib[i-1] + fib[i-2]} on Line~\ref{line:fib-overflow}; 2) the invariant
on Line~\ref{line:fib-unhold-inv} that states \CodeIn{fib[j] == fibo(j)} for all $j$
between 0 and $i$ does not hold at the end of the loop. 

\tool decides to fix the \textit{arithmetic-overflow} first. The corresponding agent
is instructed to add loop invariants or 
\CodeIn{assert}s to state the upper/lower bound of the arithmetic operations that \verus complains about. The prompt also includes tips about using 
parametric bound or
approximated bound when needed, and reasoning the monotonicity
of a data series when needed.

The output of this repair agent is highlighted by the \colorbox{green}{+2} heading
in Figure \ref{fig:repair_fib}. As we can see, this agent's 
repair includes several parts: 1) it made the right decision that
an \CodeIn{assert}, instead of a loop invariant, is needed here to state the bound on Line~\ref{line:fib-overflow-assert} (again accompanied by a comment explaining its action);
2) it figured out the right bound $0x8000\_0000$ for the 
expression \CodeIn{fib[i-1] + fib[i-2]}, probably based on the
bound expressed in the \CodeIn{fibo\_fits\_i32} spec-function in
Figure \ref{fig:fibo_spec}; 3) it realized that proving
the monotonicity of the fibonacci sequence is needed and hence
synthesized a new proof function \CodeIn{lemma\_fibo\_monotonic} 
and also used this function to support the bound assertion very nicely 
on Lines~\ref{line:fib-overflow-assert-st}--\ref{line:fib-overflow-assert-ed}.

\textit{\textbf{Iteration 3:}}
With all this effort, a simpler problem like the one in Figure \ref{fig:sum} would have been proved at this point. However, for this example, Verus now reports that the \CodeIn{while}-loop
in Figure \ref{fig:repair_fib} is completely verified, but there is a new error---the post-condition \CodeIn{fibo(n) <= fibo(m)} of 
the new function \CodeIn{lemma\_fibo\_monotonic} shown in Figure \ref{fig:repair_a} is not satisfied at the end of the function.

The agent dedicated to \textit{function-post-condition-unsatisfied} error is now called. This agent's instruction suggests adding \CodeIn{assert}s (or loop invariants) that correspond to the failed post-condition to the function exit, where the post-condition does not hold (or a relevant loop).
Therefore, the agent added the two \colorbox{green!20}{green-highlighted} lines
in Figure \ref{fig:repair_a}.
Interestingly, \revision{the \llm} does not simply assert 
\CodeIn{fibo(n) <= fibo(m)}, the failed post condition.
Instead, it added two asserts that describe the transitive
relationship of \CodeIn{fibo(n) <= fibo(m-1)}, which matches the post-condition of the proof
function right before it on Line~\ref{line:fib-lemma}, and \CodeIn{fibo(m-1) <= fibo(m)}, which is quite nice.

\textit{\textbf{Iteration 4 \& 5:}}
Unfortunately, this seemingly perfect repair is still incorrect.
Verus reports an \textit{assertion-failure} on the newly added 
Line~\ref{line:fib-lemma-fail} of
Figure \ref{fig:repair_a} --- 
\CodeIn{fibo(m-1) <= fibo(m)} cannot be verified to always hold.
Therefore, \tool calls upon its assertion-failure
repair agent. This agent is given some general options: 1) if the \CodeIn{assert}
expression is related to Verus data structure APIs like
\CodeIn{Seq::subrange}, \CodeIn{Seq::filter}, and so on, a set
of off-the-shelf lemma functions can be used; 2) change or delete
any loop invariants that might be related to the \CodeIn{assert};
3) add more \CodeIn{assert} statements; 4) add more proof
functions if needed. This fix is not easy. In fact, the first time
this repair agent is called upon, no good repair is produced.
At the second time, the repair agent comes up with the repair 
shown in Figure \ref{fig:repair_b}: it added two \CodeIn{assert}
statements to help the SMT solver reason why
\CodeIn{fibo(m-1)} cannot be larger than \CodeIn{fibo(m)}; and
it also added the \CodeIn{if m > 1} condition to avoid
negative indexing, as explained by its own comment. 

Finally, Verus reports that the whole proof task was successfully done!

What we describe above is the output from a particular run of
\tool. Since this is a difficult task, we have noticed that
\tool runs into different situations every time, and 
the final proof produced by it is not always the same.
Fortunately, with the basic proof-repair knowledge embedded in \tool agents
and the robust post-processing, \tool can reliably figure out a correct proof after a number
of attempts.

\label{sec:design}

\section{Implementation}
We implemented \tool as a command line Python tool using GPT-4o invoked by Azure OpenAI APIs. In the following, we discuss some implementation details.

\parabf{\houdini alg.} 
Given a set of proof annotations $\mathbf{A}$ that fail the verification,
if a subset of $\mathbf{A}$ is correct and sufficient to prove, \houdini algorithm
guarantees to identify this subset in a linear number of verifier invocations. 
More details of this long-established algorithm can be found in
previous work~\cite{DBLP:conf/fm/FlanaganL01}.
This algorithm has been used for GenAI-for-C
\cite{DBLP:journals/corr/abs-2311-07948} and we apply it here for \verus.

\parabf{\lynette.} To add discipline into \tool, 
we have implemented \lynette using the Verus front-end parser in Rust to post-process the LLM-generated code at the AST level. 
To check whether a \llm output $P$ is safe regarding the input Rust program and its specification $P_o$,
\lynette checks if $P$ and $P_o$ can be compiled to the same executable by comparing the pure Rust AST of the files after erasing the ghost code; \lynette compares the pre- and post-condition of $P$ and $P_o$ so the LLM cannot tweak them to change the goal of the verification; \lynette also searches for debugging ghost function, such as \CodeIn{admit()} or \CodeIn{assume()}, in $P$.  

Another use of \lynette is in program merging. Naively using a text-merging tool does not work for proof candidates, $P_1$
and $P_2$ of a
complicated proof task, as $P_1$ and $P_2$ may differ quite a lot. \lynette first erases all of the ghost code in the 
program to obtain the pure Rust AST, which is the same for $P_1$ and $P_2$, and uses it as the anchor to merge the
ghost code in $P_1$ and $P_2$.  
The minimum merging unit is a ghost expression, \ie a Verus assertion, a proof block, or an invariant.

The third use of \lynette is to support the \houdini algorithm in deleting the unprovable proof annotations from the proof candidate.
Since \lynette is AST-aware, it makes sure that the deletion honors the boundary of ghost expression, and guarantees to produce a syntactically correct result.

\parabf{Verus configuration.}
Verus has a loop-isolation configuration that can affect the difficulty of proof. When it is set to be
false, Verus will consider facts (e.g., function pre-conditions, \CodeIn{assert}s) from outside a loop body in its proof inside the loop body. By default, this configuration is set to be true for performance reasons.
If \tool fails to find a correct proof after the first two phases, it would add a \CodeIn{\#[verifier::loop\_isolation(false)]} statement to the program under proof to make proof debugging a little bit easier.

\label{sec:impl}

\section{Methodology}
\subsection{Verus-Bench}

\begin{table}
\centering
\small
\caption{Summary of \verusbench}
 \label{tab:verus-bench}
    \begin{tabular}{l|cccc|c}
\toprule
Benchmark Sources  & \clover & \diffy & \mbpp &  \misc & Total\\
\midrule
\# of Proof Tasks  & 11       & 38 & 78     & 23 & 150\\
\midrule
Executable LOC  & 175       & 951 & 1,333     & 390 & 2,849\\
Specification LOC  & 80       & 265 & 700     & 207 & 1,252\\
\bottomrule
\end{tabular}
   
\end{table}

To evaluate \tool, we need a suite of proof tasks, with each task being a Rust program annotated with
Verus specifications. 
Since such a suite does not yet exist, we crafted Verus-Bench.

We tried our best to make \verusbench a non-trivial, 
non-biased, and meaningful suite through the following methodology. First, instead of writing proof
tasks ourselves, we mainly looked into three verification-related benchmark suites in other languages and translated those tasks into \verus.
Second, we make sure \textbf{not} to include any trivial proof tasks (i.e., tasks that can be proved by Verus without any proof annotations).
In fact, 88 proof tasks from our source benchmark suites
are rejected for this reason! 
Finally, we have divided our team early so that the team who designed the \tool does not overlap with the team who set up the majority of
the \verusbench proof tasks (i.e., all those in \mbpp and \clover).

In total, \verusbench contains {150} proof tasks as shown in Table~\ref{tab:verus-bench}.

78 proof tasks are translated from
the \textbf{MBPP}-DFY-153~\cite{DBLP:journals/pacmse/MisuLM024} dataset.
That dataset contains {153} problems with specifications and verified implementation in/for Dafny, based on the \mbpp
dataset~\cite{DBLP:journals/corr/abs-2108-07732}.
Among these {153}, {67} of them require \textit{no} proof annotations to verify; {8} are 
too difficult to
translate due to code features not well supported by Rust/Verus
like floating points and string; the remaining {78} are all translated and included in \verusbench. 

11 are from
\textbf{\clover} \cite{DBLP:conf/saiv/SunSPB14}. This suite includes {60} example programs, as might be found in CS textbooks, written and verified in Dafny. The authors of \clover also provide the \verus translation for {39} of them.
Our manual inspection found that {3} of them cannot be verified by the latest 
version of Verus, {4} have weaker specifications than their variants already included in the dataset,
and {21} can be verified without proof annotations. We add all remaining {11} into \verusbench.

38 come from 
a SV-COMP-2021~\cite{SVCOMP-2021} benchmark suite, \textbf{\diffy}~\cite{DBLP:conf/cav/ChakrabortyGU20}.
\diffy contains
69 programs written in \CodeIn{C} that were designed to evaluate array and loop related
verification. Most tasks in this suite contain multiple loops 
and hence require good loop-invariant synthesis. 31 tasks
here require reasoning about non-linear equations (e.g., knowing that $(i+1)^3 == i^3 + 3*i^2 + 3*i +1$),
which is not the focus of Verus system verification. We translated all the remaining 38 tasks from $C$ to 
Rust.

The final 23 is referred to as \textbf{\misc}.
This is a collection of 23 miscellaneous Verus programs that appeared in Verus
tutorials or libraries. They include some algorithmic programs, like bubble sort, binary search, 
fibonacci sequence, and tasks that contain challenging features like nested loops, 
expressing
function post conditions as \CodeIn{assert}s instead of \CodeIn{ensures}, etc.

\par

\label{sec:bench}
\label{sec:meth}

\subsection{Experimental Setup}

\subsubsection{Hardware and software}
All our experiments are run on a machine with Ubuntu 22.04.1 LTS, 24-core CPUs, and 64 GB RAM.
By default, \tool uses GPT-4o (model version 2024-05-13). For comparison, we will also run \tool on three
other models, GPT-4-turbo, GPT-3.5-turbo, and Deepseek-R1-32B.

\subsubsection{Alternative designs in comparison} Given the huge design space of LLM-for-Verus-proof, we will refer to the following design as a \textit{baseline}. In this
design,
we repeatedly invoke GPT-4o
to generate \verus proof for the input proof task. To make this baseline competitive, we
designed its prompt to essentially be a summary of all the prompts used in \tool agents.
It uses the same \llm-temperature setting as \tool (1.0), and is configured to generate
$5$ outputs in each invocation, same as the Phase-1 agent in \tool.
Finally, we give this baseline an unfair advantage of ``cheating'': we put the answers to four
complicated tasks from \verusbench as examples in this
baseline's prompt. Notably, there is no overlap between the examples used by 
\tool agents and \verusbench.
In addition to this baseline, we will also evaluate schemes
that differ from \tool at various design points (i.e., different designs in or lack of phase 1, phase 2, and phase 3) or use different LLM models (i.e., GPT-4-turbo, GPT-3.5-turbo, Deepseek-R1) in our 
extensive ablation study. These schemes are mostly more 
competitive than the baseline above, and can offer insights about how different components of \tool work together.

\subsubsection{Evaluation metrics}
We will mainly evaluate three metrics:
\emph{number of correctly verified tasks},  
\emph{time} and  
\emph{number of \llm{} calls} required to come up with the correct proof. 
Due to the randomness of \llms, the number of correctly verified tasks may change over runs. By default, we report the number of correctly verified
 tasks after running \tool (the whole three phases) for three times; the same applies for various
variants of \tool. When comparing with the baseline, we will show the number of verified
tasks under the same time and \llm-call budget to be fair.
Keep in mind that since Verus can automatically tell whether a proof is correct, there is no manual effort
involved even if we invoke \llm or execute the \tool workflow
for multiple times. The difference between verifying a task in one attempt and in multiple attempts is
the time cost and the number of \llm-calls, which we will report.

\section{Experimental Results}
\subsection{Overall Results}

\begin{table}[t]
    \centering
    \small
    \caption{Number of tasks proved by \tool (max: 3 tries) and baseline (max: 10 min) in \verusbench.}
\begin{tabular}{lr|rrrr|r}\toprule
\multirow{2}{*}{Source} &\multirow{2}{*}{\#Tasks} & \multicolumn{4}{c|}{Proved by \tool} &\multirow{2}{*}{Proved by Baseline} \\
\cmidrule{3-6}
&    &  Total &\phaseone &\phasetwo &\phasethree &  \\\midrule
\clover &11 &11 &7 &2 &2 & 8\\
\diffy &38  &38 &26 &12 &0 & 5 \\
\mbpp &78  &68 &36 &4 &28 & 43\\
\misc &23  &20 &9 &4 &7 & 11\\ \midrule
Total &150 &137 &78 &22 &37 & 67\\
\bottomrule
\end{tabular}
    \label{tab:results}
\end{table}

Table~\ref{tab:results} shows the overall results of \tech on \verusbench.
\tech successfully proves 137 out of 150 (\emph{91.3\%}) benchmark problems. That is, provided with the
Rust implementation and specification, \tech is
able to generate proof annotations that allow Verus to prove the specification is guaranteed to hold for
137 out of 150 tasks.
Notably,  
\tech verifies all of the problems from the \clover and \diffy benchmarks.
Even though the default setting allows up to three attempts for \tool, 
\tool was able to prove 122 tasks in its first attempt, with 10 more proved in
its second attempt and 5 more on its third attempt.

As also shown in Table~\ref{tab:results},
\phaseone is able to finish slightly over half of the 150 proof tasks, while 
\phasetwo refinement and \phasethree debugging also contribute to 22 and 37 additional proven tasks respectively.
This phase-breakdown also shows differences among different benchmark sources. Since the Diffy benchmark focuses
on loop invariants, all the tasks from Diffy are handled by \tool in the first two phases. In comparison, MBPP and
Misc both heavily rely on \phasethree debugging to generate
proof annotations that go beyond loop invariants.

\revision{Processing these 137 tasks across three rounds required an average of 8.8 LLM calls per task. While approximately half of the tasks were completed with two calls or fewer, a small number required more than 40. This entire process consumed 3 million input and 1.5 million output tokens. Based on the Azure GPT-4o pricing at the time of our evaluation (\$5 per 1M input tokens and \$15 per 1M output tokens), the total cost for the benchmark suite was approximately \$37, including tasks that failed after all three rounds.}

\subsection{Comparison with Baseline}

Table \ref{tab:results} also shows that the baseline scheme of repeatedly invoking
 \llm{} (\ie{} \gpt{4o}) was only able to prove
 67 out of the 150 tasks (44.7\%), even though it was given 10 minutes for
 each task. 
Note that, this baseline scheme uses the same \llm-temperature and
$5$-output per invocation setting as \tool, and has more sophisticated prompt and examples than any \tool agent.

\begin{figure}
    \centering

    \begin{subfigure}[b]{0.5\linewidth}
        \centering
        \includegraphics[width=0.9\linewidth]{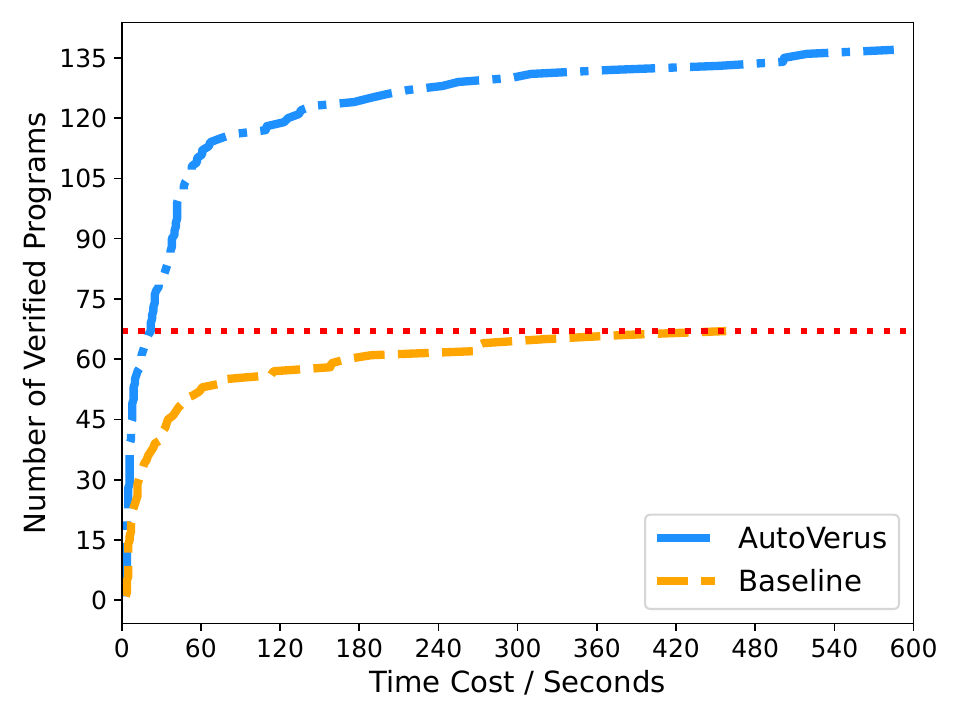}
        \caption{Tasks proved w/ different time budget.}
        \label{fig:total-time}
    \end{subfigure}%
    \hfill
    \begin{subfigure}[b]{0.5\linewidth}
        \centering
        \includegraphics[width=0.9\linewidth]{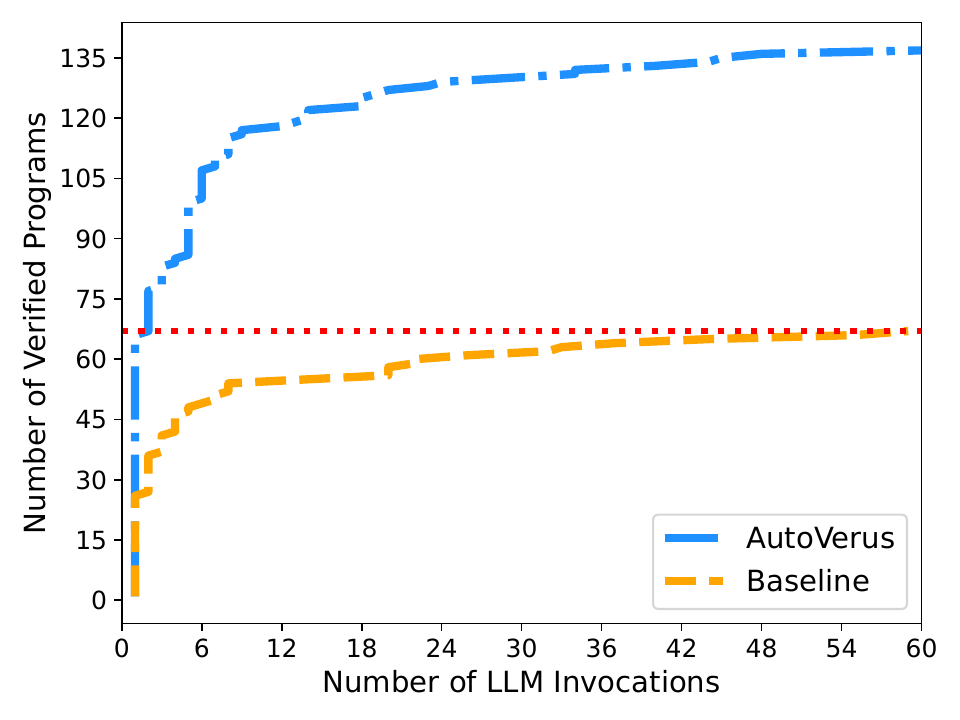}
        \caption{Tasks proved w/ \#-of-\llm-call budget.}
        \label{fig:total-invocations}
    \end{subfigure}

    \caption{Comparison with baseline in terms of time cost and \llm{} invocations.}
    \label{fig:overall-figure}
\end{figure}

The baseline does fine in generating loop invariants
for simple loops, succeeding in a similar number of tasks from
Clover, MBPP, Misc as 
Phase 1, 2 of \tool. Of course, the baseline took much
longer and many more \llm calls to achieve this than \tool, which we discuss below. 

However, the baseline cannot handle
more complicated loop invariants, and hence succeeded for only 5 out of 38 Diffy tasks. In comparison, 
\tool's Phase-1 alone generated correct proof for 26 Diffy tasks. Furthermore, the baseline scheme is poor
in generating more complicated proof annotations, and
failed many more tasks in MBPP and Misc than \tool. 

To understand how efficient \tool and the baseline are, we plotted how many
tasks can be correctly proved by each of them under a fixed time budget (up to 10 minutes per task) and the number of \llm-call
budget (up to 60 calls per task), as shown in 
Figure~\ref{fig:total-time} and Figure~\ref{fig:total-invocations}.
As we can see, \tech consistently outperforms the baseline, finishing about twice as many tasks, under the same time and \llm-call budget constraints. In order for the baseline to succeed in
67 tasks, it needs a budget of 458 seconds per task or 59 \llm-call
per task; in contrast, 22-second per-task or 2 \llm-call per
task is sufficient for \tech to generate correct proof for
at least 67 tasks. Furthermore, if we give each task only 30 seconds, \tech can generate correct proof for 81 tasks, while the
the baseline cannot even finish 40 tasks. Again, keep in mind that
the baseline has the same temperature setting and 5-output per
\llm-call as \tech.

\subsection{Detailed Results of \tool Components}

\subsubsection{The Effectiveness of Merging, Ranking, Filtering, and Houdini}

For 14 tasks, although none of the direct output of the Phase-1 agent is correct, merging some of them immediately provides correct proof just like the example in Figure \ref{fig:merge}.

During \phaseone, ranking \llm's output based on Verus output has made a difference, as more than half of the top-ranked 
proof candidates, according to Verus verification results, are \textit{not} the first output produced by the \llm{, as shown in Figure~\ref{fig:best-idx}. Interestingly, the fifth output is almost never the top-ranked candidate based on \verus, which happens to about 3\% of the model invocation.}

During the run of producing the results in Table \ref{tab:results}, \revision{\llm{s}} generated 2,637 proof candidates, of which as many as 326 (12.4\%) are unsafe.
Fortunately, they are filtered out by our \lynette tool. 

Finally, the \houdini algorithm contributed to 16 successfully proved tasks across all phases, with 5 in \phaseone, 6 in \phasetwo, and 5 in \phasethree. For each of these cases, the \houdini algorithm helped identify a correct proof from an incorrect one.

\begin{figure}
    \begin{minipage}[t]{0.46\linewidth}
\centering
    \includegraphics[width=\linewidth]{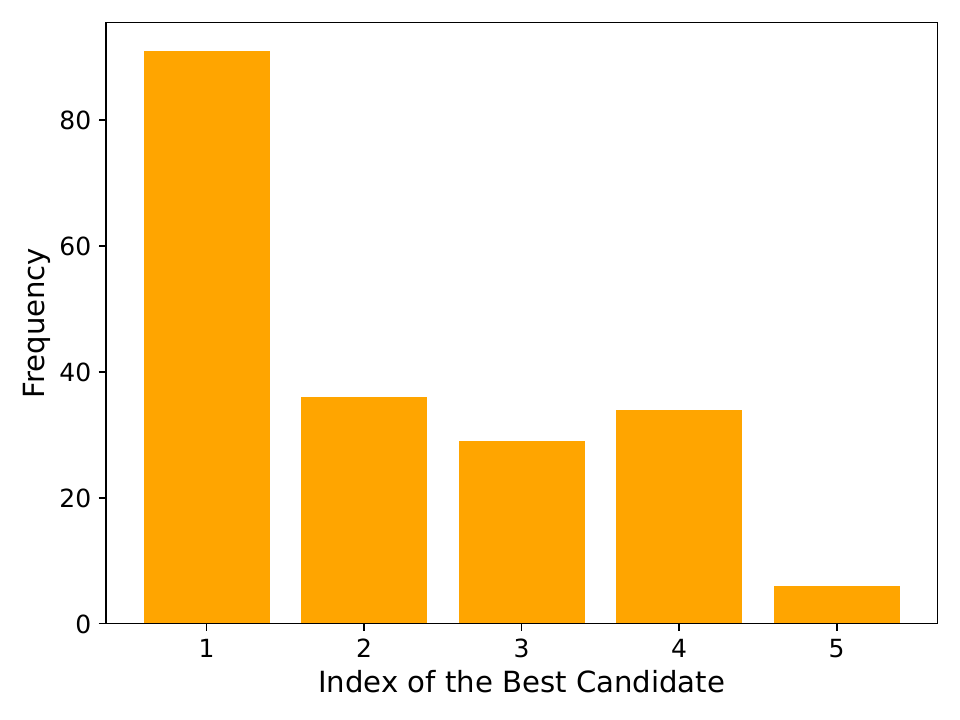}
    \caption{{Frequency of the $k$-th output of \revision{the LLM} being the best-ranked candidate according to Verus in Phase-1 (k = 1 .. 5).}}
    \label{fig:best-idx}
\end{minipage}
\hfill
\begin{minipage}[t]{0.46\linewidth}
\centering
    \includegraphics[width=\linewidth]{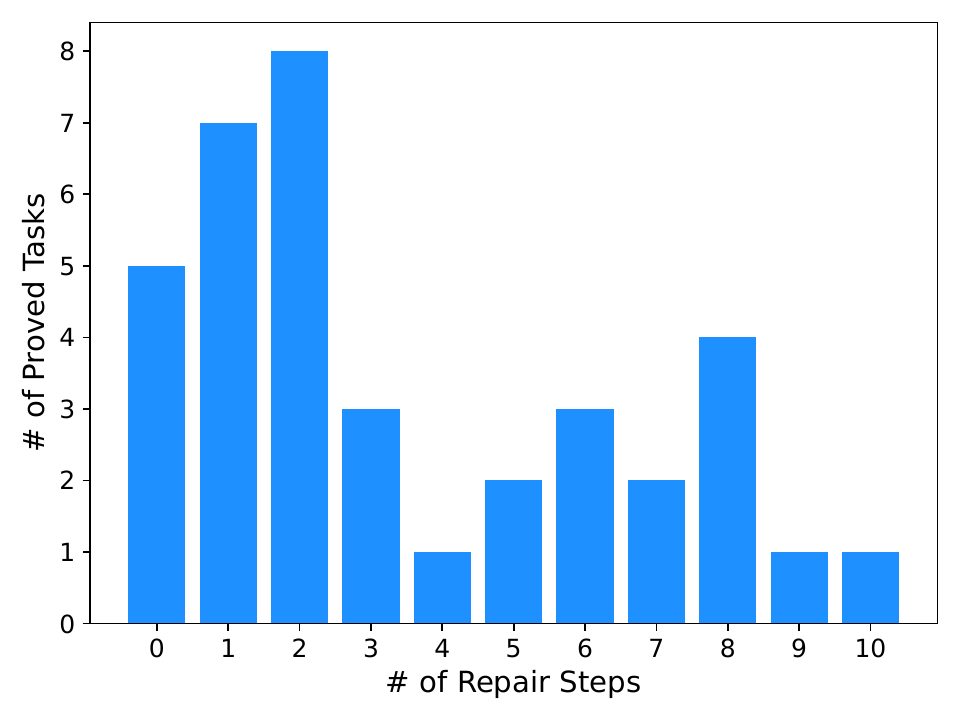}
    \caption{{Frequency of the number of debugging iterations needed for proved tasks during \phasethree.}}
    \label{fig:repair-step}
\end{minipage}

\end{figure}

\begin{table}\centering
\caption{Statistics of \tech repair agents.}
\label{tab:repair-stat}
\small
\begin{tabular}{l|r|r|rr}\toprule
Type of repair agents &\# of uses &\# of uses w/ successful repair &Repair success rate \\\midrule
\revision{AssertFail} &310 &145 &46.8\% \\
InvFailFront &131 &84 &64.1\% \\
InvFailEnd &85 &78 &91.8\% \\
ArithmeticFlow &73 &31 &42.5\% \\
PostCondFail &26 &18 &69.2\% \\
\revision{AssertFailProofFunc} &16 &11 &68.8\% \\
Default &8 &6 &75.0\% \\
PreCondVecLen &8 &4 &50.0\% \\ \midrule
Total &657 &377 &57.4\% \\
\bottomrule
\end{tabular}
\end{table}

\subsubsection{The Effectiveness of Agents}
\label{subsec:effect-agent}

For the 22 tasks that are not proved after Phase-1 but are successfully proved 
after \phasetwo (refinement) of \tool, 
6, 3, 3, and 10 tasks are proved immediately after the use of \CodeIn{Constant-Propagation}
agent, \CodeIn{Array-Length} agent, \CodeIn{Quantifier} agent, and \CodeIn{Conditional-Loop-Invariant} agent, respectively.

For the 37 tasks that require \phasethree of \tool to prove, their average number of debugging iterations
is 3.5 (median is 2). The exact distribution of debugging iterations is shown in
{Figure~\ref{fig:repair-step}.}
Only 2 tasks require more than 8 iterations to be verified.
Notably, 5 tasks are verified by adding \CodeIn{\#[verifier::loop\_isolation(false)]} to the best proof generated by \phasetwo, and hence requires 0 iteration of \llm-based repair
(as discussed in \S~\ref{sec:impl}).

\revision{To understand the contribution of individual repair agents, we analyzed their usage frequency and success rates, with a detailed breakdown shown in Table~\ref{tab:repair-stat}.}
\revision{Note, two repair agents listed in Figure \ref{fig:repair-agents} do not show up here: the ``Mismatched-Type'' repair agent operates deterministically by adding type casting without an LLM call and hence is not listed in this table; the pre-condition errors never occurred in our experiments and hence the ``PreCondFail'' repair agent was never used.}
As we can see, among all repair agents, the ones that repair Assertion Failure are used the most, for 310 times in total across 150 proof tasks, followed by the agent that repairs Loop Invariant Failed Before Loop (131 times) and at the End of the Loop (85 times). 
Due to the wide variety of assertion failures, the assertion failure agent only has a success rate of 46.8\% (i.e., in 46.8\% of times, the agent generates a repair that reduces the number of assertion failures).
In comparison, all other agents, except for the Arithmetic Overflow repair agent, have a success rate of more than 50\%. 

{Note that, only 8 out of 657 (1.2\%) repair iterations required the default repair agent. This shows that our 9 error-specific repair agents offer good coverage of the most common verification errors.}

\subsection{Ablation Study}
\label{sec:ablation}

\subsubsection{Phase-1 and \phasetwo of \tool}

As shown in 
Table~\ref{tab:ab-1}, we conducted a number of ablation studies to understand different
design choices for \phaseone and \phasetwo of \tool. For this study, we used two subsets
of proof tasks from \verusbench. The \textit{Inference Dataset} includes all the 78 tasks
that can be correctly proved by \tool using Phase-1 alone; the \textit{Refinement Dataset}
includes all the 22 tasks that require both Phase-1 and \phasetwo to be proved by \tool. 

\parabf{{Few-shot examples.}}
As shown in Table~\ref{tab:ab-1}, the number of tasks that can be
proved drops if we remove any one of the three examples used by \tool's
\phaseone agent. The third example stands out as particularly influential, resulting in the most reduction in proved tasks when removed.

\parabf{{Fusion of Phase-1 and Phase-2.}}
\revision{One may wonder the necessity of multiple separate agents in Phase-1 and Phase-2. To 
answer this question, we explored an alternative design in which all five agents in
\phaseone and \phasetwo are combined into one agent. This combined agent uses a prompt that includes the instructions from every individual
agent in the form of  ``{1. <Phase-1 Instructions>, 2. <Refinement 1 Instructions>, ...}'', and uses 
the same three few-shot examples used by the original \phaseone agent.
Under this alternative design}, 7 tasks that originally can be proved
now cannot. This includes 3 tasks that originally can be proved with just Phase-1. 
Notably, we maintain the same post-processing procedures, including filtering, ranking, merging, and the application of the \houdini algorithm, for the fused version.

\parabf{{Ranking.}}
As also shown in the table, when we discard our ranking scheme, which is based on \verus results, and instead relies on \revision{\llms} to tell us the best output, \tech verifies 12 fewer tasks.
{This demonstrates that the ranking mechanism plays a crucial role in prioritizing \llm-generated candidates effectively, leading to improved overall performance.}

\begin{table}\centering
\caption{Ablation study: \# of tasks proved under different variations of \phaseone and \phasetwo}
\label{tab:ab-1}
\small
\begin{tabular}{l|r|rrr|r|rr}\toprule
&\tech &\multicolumn{3}{c|}{Few-shot examples } &\multicolumn{1}{c|}{Fusion of} &Disable \\\cmidrule{3-5}
&default &W/o 1st &W/o 2nd &W/o 3rd &phases 1 and 2 &ranking \\\midrule
\inferenceData &78 &67 &69 &48 &75 &71 \\
\refinementData &22 & N/A & N/A & N/A &18 &17 \\
\bottomrule
\end{tabular}
\end{table}

\subsubsection{\phasethree of \tool}

We have designed an alternative generic 
debugging agent. This agent is equipped with three few-shot examples that contain examples of
\CodeIn{assert}s and proof functions, the same as what is used in our repair agents. This
agent is programmed to take in the whole Verus error report, and can decide which error
to fix and how to fix by itself.
To compare this design with the current design of \tool, we focus on those 
37 tasks that can be correctly proved by \tool through debugging. For each of these 37 tasks,
we take the imperfect proof produced by the refinement phase of \tool, which was used by
\tool's debugging phase to get the correct proof. And, we change
Verus' loop-isolation configuration to the easy-to-prove mode, 
just like that in \phasethree,
and
feed these 37 imperfect proofs to this
new debugging agent. Even with the same debugging budget (three attempts, each with up to
10 iterations), the new design only manages to produce a correct proof for 17 out of 37 tasks.

\subsubsection{\llm Choice in \tool}

\begin{table}[t]\centering
\caption{Ablation study: \tool with different \llm{s} and \llm{} temperatures.}
\label{tab:ab-sampled}
\small
\begin{tabular}{l|r|rrr|rrrr}\toprule
& \multicolumn{1}{c|}{Default} &\multicolumn{3}{c|}{\tool with other LLMs} &\multicolumn{3}{c}{Temperature} \\\cmidrule{2-8}
& GPT-4o, temp=1.0 &GPT-4-turbo &GPT-3.5-turbo & Deepseek-R1&0.1 &0.4 &0.7 \\\midrule
\sampleData &26 &26 &18 & 23 &21 &24 &25 \\
\bottomrule
\end{tabular}
\end{table}

\begin{figure}
    \centering
    \includegraphics[width=0.8\linewidth]{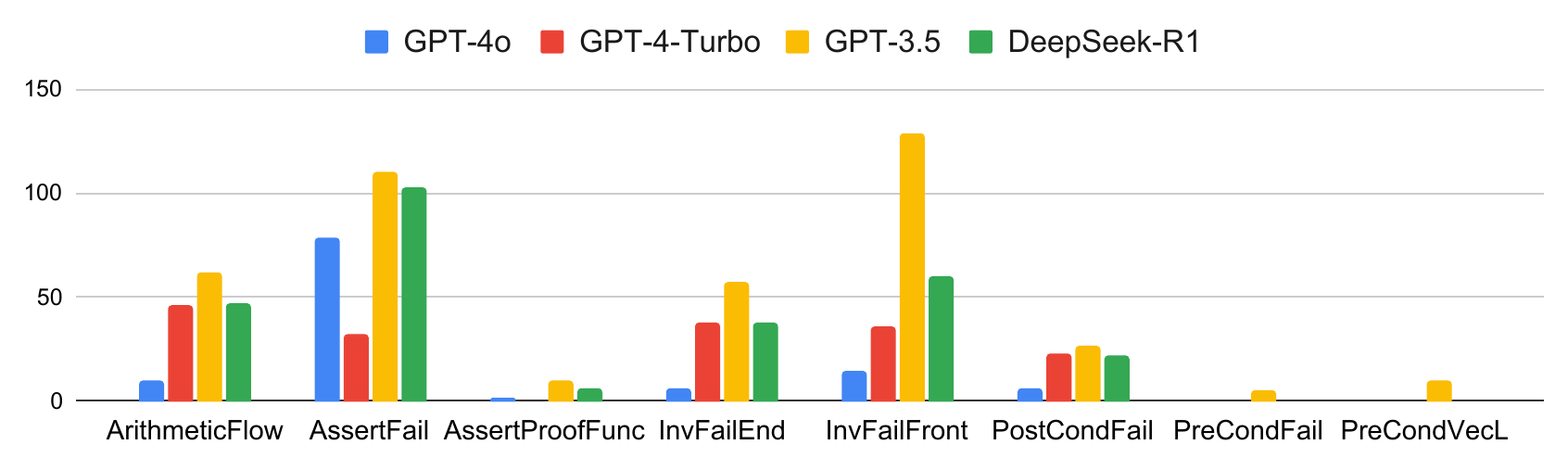}
    \caption{\revision{The number of repair-agent invocations by \tool with different \llm{s} on the 30 sampled tasks.}}
    \label{fig:ab-repair-agent}
\end{figure}

Due to resource constraints, instead of using the whole \verusbench, we conduct this part of the ablation study using 
a smaller dataset of 30 problems randomly sampled from
\verusbench, preserving the
original benchmark-source distribution (i.e., 2 from \clover, 8 \diffy, 15 \mbpp, and 5 \misc).

Our default setting (with GPT-4o, temperature 1.0) proves 26 tasks, the most among all settings (shown in Table~\ref{tab:ab-sampled}). Note, the baseline scheme of repeatedly invoking GPT-4o can only prove 14 tasks.
While trying three different GPT models, 
we observe that \gpt{4-turbo} delivers competitive performance compared to our default setting with \gpt{4o}, with both models power \tool to prove 26 out of the 30 tasks. In contrast, \gpt{3.5-turbo}
only allows \tool to prove 18 tasks. However,
upon further analysis, we find that \gpt{4-turbo} 
requires significantly more time in \tool to figure out the correct proofx---an average of 229.8 seconds per task, compared to just 70.0 seconds for \tool with \gpt{4o}.  
{Furthermore, benefiting from the \tool design, \gpt{3.5-turbo} outperforms the baseline of repeatedly invoking GPT-4o, proving 18 tasks. 

We have also evaluated \tool with the open-source model Deepseek-R1-32B, and have observed a slightly worse performance than \tool with GPT-4 models (23 vs. 26). It shows that \tool can work with different types of models. It also shows that code-proof remains a challenging task even for the new generation of reasoning models.

{Figure~\ref{fig:ab-repair-agent} presents the number of repair-agent invocations by \tool with different underlying models for these 30 sampled tasks.
\revision{The figure omits the non-LLM ``Mismatched-Type'' agent and the generic repair agent (``repair\_default''), which was never triggered in this ablation experiment.}
Generally speaking, \tool invokes a similar set of repair agents with different \llm{s}: the repair agent for assertion failures is most frequently used for all models except for GPT-4-Turbo; the popularity of repair agents for arithmetic under/overflow errors, two types of invariant failures,
and post-condition failures are similar; the two types of pre-condition failures are rarely encountered during proof debugging across all models. Of course, the exact number of invocations are very different across different \llm{s}: \tool has to make the most repair-agent invocations with GPT-3.5, probably because the raw output of GPT-3.5 has the worst quality among all models; on the other hand, \tool needs the fewest repair-agent invocations with GPT-4o.
Overall, this result shows that the design of \tool is generally applicable for different \llm models, and can effectively help boost the performance of \llm{s}.}

Finally, while setting the temperatures of \gpt{4o} to be
0.1, 0.4, 0.7, and 1.0, \tool proves 
21, 24,   25,  and 26 tasks, respectively.
This aligns with our design intuition that 
a higher temperature setting is beneficial in fostering 
\llm's creativity and exploring a wider range of potential solutions,
as long as human experts and discipline are also applied.

\subsection{\revision{Looking into \tool Proofs}}

\parabf{Failed cases.}
\revision{\tech failed to prove 13 tasks. We categorize \tool's struggle into three types.
First, \tool sometimes struggles at proving an expression will never encounter arithmetic overflow/underflow.
For example, task \CodeIn{MBPP-170} contains a loop that iterates through every element of an array to compute their sum. Verus requires a proof that the statement ``\CodeIn{sum = sum + arr[index] as i128}''
inside the loop body would never encounter arithmetic overflow, where \CodeIn{arr} has the type of \CodeIn{Vec<i64>}, 
\CodeIn{index} has the type of \CodeIn{u64}, and \CodeIn{sum} has the type of \CodeIn{i128}. In the human-provided
proof, this is achieved through a bound approximation: ``\CodeIn{sum <= i64::MAX * index}'' is added as a loop invariant,
which nicely provides a bound for \CodeIn{sum}. In \tool, unfortunately, \llm tried to directly prove ``\CodeIn{sum + arr[index] <= i128::MAX}'', and failed. In total, this type of challenge caused \tool to fail 6 tasks.}

\revision{Second, Verus has a \CodeIn{vstd} library about proof-specific data structures (\eg{} \CodeIn{Set}, \CodeIn{Seq}, \CodeIn{Map}), where axioms and helper
lemma functions are provided. Without full knowledge of the \CodeIn{vstd}, \tool struggles at proof
tasks that involve complicated usage of Verus data structures. For example, the post-conditions in the
\CodeIn{Misc-deduplicate} task involve two \CodeIn{vstd} APIs, 
\CodeIn{Seq::no\_duplicates()} and \CodeIn{Seq::to\_set()}. 
Without knowing the exact definitions and axioms related to these two APIs,
proof writing is difficult and \tool failed. In total, this led to the failure in 5 tasks.}

\revision{Finally, \tool seemed lost and failed on 2 tasks with complicated logic. For example, the proof task of \CodeIn{MBPP-755} involves a recursive spec-function and two complicated function post-conditions as shown below. The proof attempt of \tool was quite far from success.}
\begin{lstlisting}[language=Verus]
ensures
  forall|k: int| 0 <= k < numbers.len() && k != indices.0 && numbers[indices.0] == min_spec(numbers@) ==> (#[trigger] numbers[k] >= numbers[indices.1]),
  exists|k: int| 0 <= k < numbers.len() && k != indices.0
      && (#[trigger] numbers[k] == numbers[indices.1])
\end{lstlisting}

\parabf{Successful cases.}
\revision{For the 137 tasks that \tool successfully proved, at a high level, the quality
of \tool proof is good and in line with human-written proof --- after all, passing the
Verus verification is a high bar for quality. }

\begin{wrapfigure}{r}{0.485\linewidth} %
    \centering
    \includegraphics[width=\linewidth]{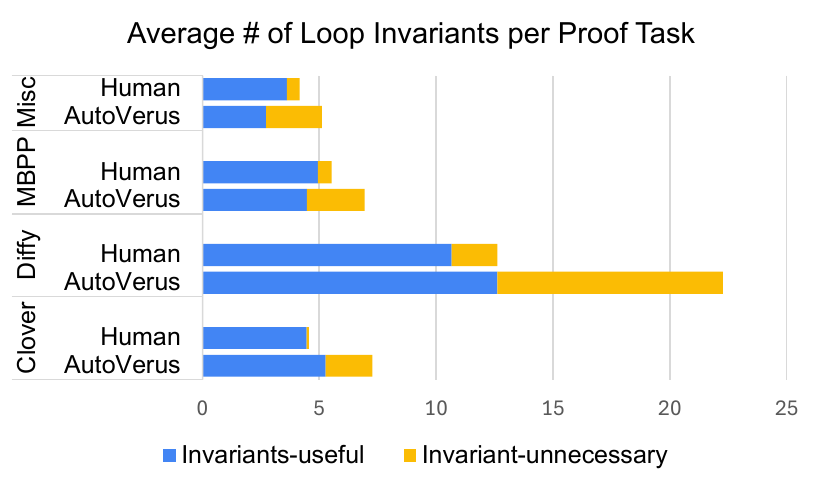}
    \caption{\revision{Average number of necessary vs. unnecessary loop invariants from \tool and human authors, evaluated on tasks successfully proven by \tool.}}
    \label{fig:proof_stats}
\end{wrapfigure} %
\revision{To quantitatively compare \tool proof and
human proof, we calculated the average number of loop invariants\footnote{The proof annotations include not only loop invariants, but also \texttt{assert} statements and lemma functions. However,
since \texttt{assert} and lemma functions are not needed by some benchmarks (e.g., all the Diffy benchmarks), we focus our study here on loop invariants.} 
across these tasks
that \tool succeeded on and also identified which of these invariants are necessary.
While every loop invariant in a valid proof is correct, it may not be needed for the ultimate goal of a proof task (\ie function post-conditions).
To distinguish necessary invariants from unnecessary ones, we have designed a tool that automatically removes one loop invariant at a time.
If \verus can still verify the program after the removal, this loop invariant is considered unnecessary and discarded. Otherwise, it is useful and restored. This process repeats until no loop invariant can be removed without causing the proof to fail.
We performed the same analysis on human-written proofs for comparison, as shown in Figure~\ref{fig:proof_stats}.}

\revision{As we can see in the blue bars of Figure \ref{fig:proof_stats}, proofs generated by \tool and those by humans contain a similar number of useful loop invariants. \tool used slightly fewer useful loop invariants on Misc and MBPP tasks, while the trend is reversed on Diffy and Clover tasks.}

\revision{As we can also see in the orange bars of Figure~\ref{fig:proof_stats}, a more significant difference between \tool and human proofs lies in the number of unnecessary loop invariants, which are far more prevalent in \tool proofs. For example, on the Diffy benchmark, \tool generated an average of 22.2 loop invariants per task, 9.6 of which were unnecessary. In contrast, human proofs for the same tasks contained 12.6 total invariants, with only 2.0 being unnecessary. These unnecessary loop invariants 
exist even in human-written proofs, as it is challenging to identify the minimal set of evidence required by the verifier to finish a proof. For \tool, this issue is exacerbated because \llm agents sometimes add identical loop invariants across multiple proof attempts. While this verbosity can harm code and proof readability, the negative impact is limited, as an automated process can effectively remove these unnecessary annotations.}

\label{sec:results}

\section{Discussion \& Threats to Validity}
\parabf{General threats to validity.}
There is an inherent randomness in \tool and our evaluation given the probabilistic nature
of \llm. We have made our best effort to make \verusbench an unbiased and meaningful benchmark suite.
However, it is still just a collection of 150 tasks. It cannot represent
many real-world proof tasks in large system code; it also cannot guarantee to represent all small-scale
proof tasks. We have tried our best not to tune \tool based on features of specific proof tasks
in our \verusbench, and have our team that sets up the majority of the \verusbench not participating
in the design of \tool. However, the authors who participated in the design of \tool are already
familiar with some of the tasks in the Diffy benchmark and in the Misc set. Finally, 
our evaluation does not yet cover some of the newest Rust features supported by \verus, such as iterators.

\parabf{Data leakage.}
\revision{Given the huge amount of training data used by modern \llms, one may wonder whether \llms already saw some of the proof tasks in its training data. We believe the impact of potential data leakage is minimal for two reasons:}

\revision{
First, the (un-)availability of online Verus resources. Among the four sources of \verusbench, Verus proofs for all 38 Diffy tasks and all 78 MBPP tasks were not publicly available throughout this research project. 
The 11 examples from
CloverBench were released in the late spring of 2024, while we utilized earlier GPT-4o models (with \CodeIn{2024-02-15} and \CodeIn{2024-05-13} versions), predating the widespread public availability of CloverBench.
The only concerns are three tasks in the Misc part of \verusbench, ``fibonacci'', ``tail\_triangle'', and ``lemma\_len\_intersect'': these three tasks are created based on examples in the online
Verus Tutorial \cite{verustutorial}. We will elaborate on these three cases next.}

\revision{
Second, the poor performance of baseline \llm on ``leaked'' tasks. Without the support of \tool, the baseline \llm (\gpt{4o}) is unable to generate proof for the three tasks that are already explained in the online Verus Tutorial.
\tool was able to conquer these three tasks. However, ``fibonacci'' remains one of the most difficult tasks even for \tool as
discussed in \ref{sec:fibo}.
Plus, the final proof generated by \tool is
{\it different} from the proof offered in Verus Tutorial.
Comparing Figure \ref{fig:repair_b} generated by \tool and the listing below from the Verus Tutorial, we can see that although both functions prove the monotonicity of the Fibonacci sequence, they divide the proof into different sets of cases and also conduct the inductive proof differently. Based on these results, we suspect that existing Verus online resources are too few for LLMs to remember or understand.}

\begin{lstlisting}[language=Verus, escapechar=!]
proof fn lemma_fib_is_monotonic(i: nat, j: nat)
    requires i <= j, ensures fib(i) <= fib(j), decreases j - i
{
    if j < 2 { } else if i == j { } else if i == j - 1 { } else {
        lemma_fib_is_monotonic(i, (j - 1) as nat);
        lemma_fib_is_monotonic(i, (j - 2) as nat);
    }
}
\end{lstlisting}

\parabf{How to scale up?}
Moving forward, we expect three key challenges in applying \tool to real-world project-level system verification. The first challenge comes from much more complicated code dependencies in large systems. Verus allows code proof to be conducted at the function level, which helps address part of the scalability concern of \tool. However, a function in a large system often calls other functions, both executable functions and specification functions in the world of Verus. Tackling such code dependency may require a hybrid approach of GenAI and traditional program analysis.
The second challenge is related to specification. \tool assumes that the specification (i.e., function pre-conditions and post-conditions) is already given. In a large system, deciding the specification of one function could be part of the proof task for another function. Furthermore, the quality of specification can greatly affect the proof task.
\revision{The final challenge, and maybe the most difficult one, is to handle more complicated and diverse proof problems. In real-world verified systems \cite{verifiedstorage, verussosp24, DBLP:conf/osdi/ZhouACGHC24}, each function is often associated with a long list of pre-conditions and a long list of post-conditions, whose complexity is much higher than that in \verusbench. The proof skills used in those systems are also much more diverse --- indeed, many proof annotations in those systems use advanced features of Verus \cite{verustutorial} that are not covered by \tool yet.}
We believe all three challenges will be interesting for future research to explore.

\section{Related Work}

\parabf{Verification for Rust.}
As the popularity of Rust continues to grow, a variety of tools have been developed to verify Rust programs~\cite{DBLP:conf/icfem/DenisJM22, 
DBLP:journals/pacmpl/Astrauskas0PS19, DBLP:journals/pacmpl/HoP22, 
DBLP:journals/pacmpl/0002JKD18, DBLP:conf/pldi/MatsushitaDJD22, DBLP:journals/pacmpl/LattuadaHCBSZHPH23, DBLP:journals/pacmpl/GaherSJKD24, kani2024}. 
Among these, Verus and Creusot stand out by allowing programmers to write specifications and proofs directly in Rust.
Verus distinguishes itself further by integrating Rust’s borrow checker into its verification process, enforcing linearity and borrowing in specifications and proofs. 
This design simplifies reasoning about pointers and concurrency, making Verus particularly suited for systems verification. 
\cite{DBLP:conf/osdi/0013MGMCH0PSSX24, DBLP:conf/osdi/ZhouACGHC24}.

\parabf{LLMs for proof synthesis.}
\llm{s} have been revolutionizing various software engineering tasks~\cite{xia2023keep,tian2024debugbench,deng2023large,yang2023white,yang2025kernelgpt}, especially for code generation~\cite{chen2021evaluating,DBLP:journals/corr/abs-2108-07732},
but the reliability of the generated code remains uncertain~\cite{DBLP:conf/nips/LiuXW023}. 
Formal proofs offer mathematical guarantees of correctness against program specifications, yet writing these proofs remains a labor-intensive, expertise-driven process.
This challenge has sparked interest in using LLMs to automate formal proof synthesis ~\cite{DBLP:journals/corr/abs-2404-09939}.

Before LLMs, researchers explored using neural networks to synthesize proof for interactive theorem provers such as Coq~\cite{coq}, Isabelle/HOL~\cite{isabelle}, and LEAN~\cite{DBLP:conf/cade/Moura021}, where proof to a theorem is completed by a sequence of proof steps, each being a proof tactic with its arguments \cite{DBLP:conf/icml/YangD19, bansal2019holist, DBLP:journals/pacmpl/FirstBG20, Proverbot9001, DBLP:conf/icse/FirstB22, polu2020generative, DBLP:conf/nips/JiangLTCOMWJ22}. 
Recent efforts like LeanDojo~\cite{DBLP:conf/nips/YangSGCSYGPA23}, LEGO-Prover~\cite{DBLP:conf/iclr/WangXZLCHXSX0LL24}, and Baldur~\cite{DBLP:conf/sigsoft/FirstRRB23} have used \llms, and they all leverage the vast existing
data of interactive theorem provers to create RAG databases or fine-tune models.
However, they only work for interactive theorem provers and are primarily designed for maths rather than programs.
Although it is theoretically possible to transpile programs in Rust to interactive theorem provers and then apply those tools, the cost is impractically high.

Recent work also tried using \llms to generate proof/code
in proof-oriented languages like Dafny~\cite{DBLP:conf/saiv/SunSPB14,DBLP:journals/pacmse/MisuLM024, DBLP:journals/corr/abs-2405-16792,DBLP:journals/corr/abs-2406-08467} 
and F$^*$~\cite{DBLP:journals/corr/abs-2405-01787}.
As discussed in \S~\ref{sec:back}, implementation and proof are
intertwined in F$^*$, and hence the proof generation faces different challenges
from \tech. Recent work~\cite{DBLP:journals/corr/abs-2405-01787} leverages
the large quantity of existing F$^*$ code to create fine-tuned small models
and a RAG database that is used to augment its proof synthesis prompt.

Dafny is also an SMT-based verifier like \verus. It proves programs written in the Dafny language.
None of the LLM-for-Dafny work focuses on synthesizing proof for a given code implementation.
Thus, their designs and evaluation results are very different from \tech.
Laurel~\cite{DBLP:journals/corr/abs-2405-16792} aims to 
synthesize \CodeIn{assert}s to verify an existing proof function.
It uses a lemma-similarity score to help \revision{\llms} learn from similar \CodeIn{assert} in the same code base, with a success rate of over 50\%.
It could help the Assertion Failure repair agent in \tech in the context of a mostly verified code base.
Two recent projects \cite{DBLP:conf/saiv/SunSPB14, DBLP:journals/pacmse/MisuLM024} aim to create 
a workflow from natural language description to verified Dafny programs.
Since their focus was not on generating proof annotations for a given
implementation, they simply feed the \llm with one generic prompt.
Given the difficulty of generating the whole program based on natural language, most of the resulting Dafny programs contain easy or no proof annotations, as discussed in \S~\ref{sec:bench}.
DafnyBench \cite{DBLP:journals/corr/abs-2406-08467} creates a dataset of more than 750 Dafny programs and compares the proof-synthesis capability across different \llms. 
It feeds the \llm with the program under proof and one generic prompt. When the \llm's initial answer
is wrong, DafnyBench adds the whole error message into the prompt, similar to the alternative design of \tool's debugging agent studied in \S~\ref{sec:ablation}.
Note, 26.9\% of the proof tasks in DafnyBench do not require any proof annotations.
This type of tasks is not included in \verusbench.

Lemur \cite{DBLP:conf/iclr/0001BN24} transforms program verification tasks into a series of deductive steps suggested
by \llm{s} (in the form of program invariants) and subsequently validated by automated reasoners (a model checker in 
Lemur). The current Lemur tool cannot be directly applied to Verus (\eg the type of program invariants and benchmark problems discussed there do not involve quantifiers or proof functions). However, the high-level idea could be beneficial for future GenAI-for-Verus.

Recently, Loopy uses \llm to generate loop invariants in integer C programs (\ie no arrays or pointers) \cite{DBLP:journals/corr/abs-2311-07948}.
It uses a long prompt to generate loop invariants, and uses Houdini algorithm to identify the correct subset of loop invariants.
If Houdini fails, it feeds the whole error message to a long fixed repair prompt.
\tool shares the similarity with Loopy as both leveraging Houdini algorithm.
Loopy only focuses on loop invariants that do \textit{not} involve quantifiers, arrays, etc., while \tool does not have these constraints for loop-invariant synthesis. More importantly, as discussed earlier, \tool is much more than just synthesizing loop invariants. In fact, among the 150 Verus-Bench proof tasks, only the 38 Diffy tasks focus on loop invariants. The majority of the proof tasks in the other three sources (Clover, MBPP, Misc) cannot be proved by adding loop invariants alone--for example, 43 out of 78 MBPP tasks and 11 out of 23 Misc tasks \emph{cannot} be proved using loop invariants alone.
Due to the wide variety of proof annotations
that \tool needs to deal with (\CodeIn{assert}, proof functions, 
loop invariants, quantifiers, collection types, etc.), its design is completely
different from Loopy.

\section{Conclusions}
In this paper, we explored using GenAI techniques to generate Verus proof for Rust code. 
By embracing the unique features of Verus, we have designed \tool that uses human expert 
knowledge and Verus-based formal methods to assist \revision{\llms} in writing proof. \tool has achieved
 good results on \verusbench, and provides a starting point for future research.

\revision{The \llm} is quickly evolving and \verus is also evolving. It is possible that some of 
the designs in \tool will need to change when the next big \llm is released, just like almost
every AI-based technique these days. We have designed \tool to contain composable parts
to ease future extension and improvement. We also believe some of the key insights in \tool
will last for a long time, such as the power of
the combined capability of human expertise, \llm creativity, and formal methods' rigor.
We also hope our exploration to accommodate various features of Verus will be an 
interesting case study for other AI-for-software efforts.

\section*{Acknowledgment}
We are grateful to the anonymous reviewers for their valuable comments and feedback.

\section*{Data-Availability Statement}
The code and artifact of \tech are available at \href{https://github.com/microsoft/verus-proof-synthesis}{microsoft/verus-proof-synthesis}.

\bibliographystyle{ACM-Reference-Format}
\bibliography{main}

\end{document}